\documentclass[aps, prb, amsfonts, amssymb, amsmath, reprint, showkeys, nofootinbib, twoside, superscriptaddress]{revtex4-2}
\usepackage{mathbbol}
\usepackage[english]{babel}
\usepackage[utf8]{inputenc}
\usepackage[colorinlistoftodos, color=green!40, prependcaption]{todonotes}
\usepackage{comment}
\usepackage{array}

\usepackage[acronym]{glossaries}
\usepackage{amsthm}
\usepackage{mathtools}
\usepackage{physics}
\usepackage{xcolor}
\usepackage{graphicx}
\usepackage[left=23mm,right=13mm,top=35mm,columnsep=15pt]{geometry} 
\usepackage{adjustbox}
\usepackage{placeins}
\usepackage[T1]{fontenc}
\usepackage{lipsum}
\usepackage{csquotes}
\usepackage{multirow}
\usepackage[cal=boondoxo]{mathalfa}

\usepackage[pdftex, pdftitle={Article}, pdfauthor={Ignacio Loaiza}]{hyperref} 
\usepackage{rotating}
\usepackage{xcolor}

\newcolumntype{P}[1]{>{\centering\arraybackslash}p{#1}}
\newcolumntype{M}[1]{>{\centering\arraybackslash}m{#1}}

\newcommand{\prep}[0]{$\mathtt{PREPARE}$ }
\newcommand{\sel}[0]{$\mathtt{SELECT}$ }

\begin{document}
\title{Majorana Tensor Decomposition: A unifying framework for decompositions of fermionic Hamiltonians to Linear Combination of Unitaries}

\author{Ignacio Loaiza}
\affiliation{University of Toronto, Department of Chemistry, Chemical Physics Theory Group, Toronto, ON, Canada}
\affiliation{University of Toronto Scarborough, Department of Physical and Environmental Sciences, Toronto, ON, Canada}
\author{Aritra Sankar Brahmachari}
\affiliation{Indian Institute of Science Education and Research - Kolkata}
\author{Artur F. Izmaylov}
\affiliation{University of Toronto, Department of Chemistry, Chemical Physics Theory Group, Toronto, ON, Canada}
\affiliation{University of Toronto Scarborough, Department of Physical and Environmental Sciences, Toronto, ON, Canada}

\date{\today} 

\newacronym{LCU}{LCU}{linear combination of unitaries}
\newacronym{QPE}{QPE}{quantum phase estimation}
\newacronym{MTD}{MTD}{Majorana tensor decomposition}
\newacronym{SVD}{SVD}{singular value decomposition}
\newacronym{MPS}{MPS}{matrix product state}
\newacronym{SF}{SF}{single factorization}
\newacronym{DF}{DF}{double factorization}
\newacronym{AC}{AC}{anti-commuting}
\newacronym{CSA}{CSA}{Cartan sub-algebra}
\newacronym{THC}{THC}{tensor hypercontraction}
\newcommand{\myceil}[1]{\lceil #1 \rceil}
\newcommand{\myfloor}[1]{\lfloor #1 \rfloor}
\newcommand{\bea}{\begin{eqnarray}}
\newcommand{\eea}{\end{eqnarray}}
\newcommand{\EQ}[1]{Equation~(\ref{#1})} %
\newcommand{\eq}[1]{Eq.~(\ref{#1})} %
\newcommand{\eqs}[1]{Eqs.~(\ref{#1})} %
\newcommand{\fig}[1]{Fig.~\ref{#1}} %
\newcommand{\figs}[1]{Figs.~\ref{#1}} %

\begin{abstract}
\Gls{LCU} decompositions have appeared as one of the main tools for encoding operators on quantum computers, allowing efficient implementations of arbitrary operators. In particular, \gls{LCU} approaches present a way of encoding information from the electronic structure Hamiltonian into a quantum circuit. Over the past years, many different decomposition techniques have appeared for the electronic structure Hamiltonian. Here we present the Majorana Tensor Decomposition (MTD), a framework that unifies existing \glspl{LCU} and offers novel decomposition methods by using popular low-rank tensor factorizations.
\end{abstract}


\maketitle

\section{Introduction}
Quantum chemistry has appeared as one of the main contenders in the race for quantum supremacy. This problem consists in finding eigenvalues of the electronic structure Hamiltonian
\begin{equation} \label{eq:el_ham}
    \hat H = \sum_{ij}^N h_{ij} \hat F^i_j + \sum_{ijkl}^N g_{ijkl} \hat F^i_j \hat F^k_l,
\end{equation}
where $\{i,j,k,l\}$ are spacial orbitals, $h_{ij}$ and $g_{ijkl}$ are one- and two-electron integrals\footnote{Representing the Hamiltonian using only excitation operators is usually referred to as chemists' notation. This entails a modification to the one-electron tensor with respect to physicists' notation, which uses normal-ordered operators of the form $\hat a^\dagger_p\hat a^\dagger_q \hat a_r \hat a_s$. Our notation is related to the electronic integrals by $g_{ijkl} = \frac{1}{2}\int \int d\vec r_1 d\vec r_2 \frac{\phi_i^*(\vec r_1)\phi_j(\vec r_1)\phi_k^*(\vec r_2)\phi_l(\vec r_2)}{|\vec r_1 - \vec r_2|}$ and $h_{ij} =- \sum_k g_{ikkj} + \int d\vec r \phi_i^*(\vec r) \Big(-\frac{\nabla^2}{2} - \sum_n \frac{Z_n}{|\vec r - \vec R_n|} \Big) \phi_j(\vec r)$, with $\phi_i(\vec r)$ the one-particle electronic basis functions, and $Z_n/\vec R_n$ the charge/position of nucleus $n$.}, $N$ is the number of spacial one-electron orbitals, and $\hat F^i_j \equiv \sum_{\sigma} \hat a^\dagger_{i\sigma}\hat a_{j\sigma}$ are spacial excitation operators, with $\sigma\in\{\alpha,\beta\}$ spin-z projections.  \\

\begin{figure}
    \centering
    \includegraphics[width=8.5cm]{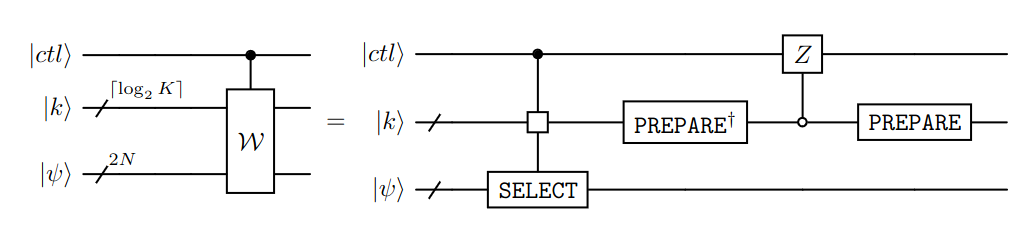}
    \caption{LCU-based implementation of walk operator with eigenvalues $e^{\pm i \arccos\frac{\hat H}{\lambda}}$, for $\lambda = \sum_k |c_k|$ the 1-norm of the LCU. Note the Z gate is applied conditioned on the state of all qubits in register $k$ to be zero.}
    \label{fig:walk}
\end{figure}

Efficiently finding eigenvalues of the electronic structure Hamiltonian can be achieved on a quantum computer through algorithms such as \gls{QPE} and its variants \cite{QPE,ts_qpe,cirac_qpe,aspuru_qpe,heisenberg_qpe,statistical_qpe,kitaev_QPE,qubitized_pe,GSEE}. These algorithms require the implementation of functions of the Hamiltonian, such as the time evolution operator $e^{-i\hat{H}t}$, within the quantum circuit to incorporate information from $\hat{H}$. Generally, there are two approaches for encoding $\hat{H}$ on a quantum circuit: Trotter product formulas \cite{product_1,product_2,product_xanadu} and \gls{LCU}-based encodings. Product formula approaches involve decomposing $\hat{H}$ into a sum of fast-forwardable fragments to implement the time-evolution operator. Despite its advantages and low qubit costs, this family of approaches does not provide the optimal $\sim \mathcal{O}(t)$ time scaling \cite{trotter,childs2021theory,trotter_step}. On the other hand, \gls{LCU}-based encodings construct an $\hat{H}$ oracle circuit and enable the implementation of an arbitrary polynomial of $\hat{H}$ \cite{qsvt,low2019hamiltonian,gqsp}. The LCU encoding also gives rise to the qubitization approach implementing 
a walk operator \cite{walk,low2019hamiltonian,THC} denoted as $\hat{\mathcal{W}}$, which has eigenvalues $e^{\pm i \textrm{arccos} (E_n/\lambda)}$, where $E_n$ are the eigenvalues of $\hat{H}$ (see Fig.\ref{fig:walk}) and $\lambda=\sum_k |u_k|$ is a 1-norm of a coefficient vector for an LCU decomposition of $H$
\begin{equation} \label{eq:lcu}
    \hat{H} = \sum_{k=1}^K u_k \hat{U}_k.
\end{equation}
Here, $\hat{U}_k$ are unitary operators, $u_k$ are complex coefficients, and $K$ is the total number of unitaries.
By estimating the phase of the walk operator, the eigenvalues of $\hat H$ can be extracted from $\textrm{arccos}(E_n/\lambda)$. The accuracy associated with the energy estimation is upper-bounded by $\epsilon \leq \lambda \epsilon_W$, where $\epsilon_W$ is the walk operator phase estimation accuracy. Consequently, employing an optimal phase estimation algorithm with Heisenberg scaling $\tilde{\mathcal{O}}(1/\epsilon_W)$ \cite{heisenberg_qpe,GSEE}, the algorithm's cost to achieve a target energy accuracy of $\epsilon$ would require $\tilde{\mathcal{O}}(\lambda/\epsilon)$ calls to the walk operator. It is evident that the cost of the LCU-based energy estimation method depends on both the 1-norm $\lambda$ and the implementation cost of the walk operator. 

\begin{figure}
    \centering
    \includegraphics[width=8cm]{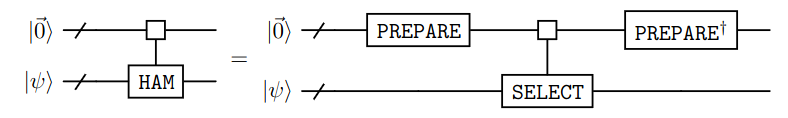}
    \caption{LCU Hamiltonian oracle implementation by block encoding with output $\ket{0}_a\otimes\frac{\hat H}{\lambda}\ket{\psi} + \ket{\textrm{remainder}}$. \prep$\ket{0}_a = \sum_k \sqrt{\frac{u_k}{\lambda}}\ket{k}_a$ and \sel $= \sum_k \ket{k}\bra{k}_a \otimes \hat U_k$ for the \gls{LCU} shown in Eq.~\eqref{eq:lcu} and $\ket{k}_a$ the $k$-th state of the ancilla register, requiring $\myceil{\log_2 M}$ ancilla qubits, where $M$ is the total number of unitaries in the \gls{LCU}.}
    \label{fig:lcu_oracle}
\end{figure}

In this work, we consider lowering the circuit cost of the LCU-based Hamiltonian encoding \cite{LCU} by optimizing the choice of unitary operators. The circuit cost of LCU decomposition is proportional to its 1-norm $\lambda$. By employing the LCU decomposition, we can construct a Hamiltonian oracle, which corresponds to a circuit that encodes the action of $\hat{H}/\lambda$ on a quantum register \cite{interaction,LCU}, as illustrated in Fig.~\ref{fig:lcu_oracle}. The spectral range $\Delta E \equiv E_{\max} - E_{\min}$, where $E_{\max(\min)}$ represents the maximum (minimum) eigenvalue of $\hat{H}$, provides a lower bound for the 1-norm, given by $\lambda \geq \Delta E/2$ \cite{loaiza_lcu}. Based on this consideration, it becomes evident that the eigenvalues of $\hat{H}/\lambda$ after a constant shift will fall within the range of $[-1,1]$\footnote{A constant shift should also be considered to center the spectrum around $0$. However, the discussion of such constant shifts is omitted as they can be easily added due to their commutativity with all operators.}. Having the Hamiltonian's spectrum within this range is necessary to ensure that the block encoding defines a unitary transformation in the quantum register plus ancilla space.
Numerous LCU decompositions have been proposed for the electronic structure Hamiltonian \cite{femoco_df,THC,loaiza_lcu,bliss}. 
In this work, we present a unified framework for expressing the electronic structure Hamiltonian as an LCU. In addition, by working using polynomials of Majorana operators, we expect these decompositions to be applicable to other kind of Hamiltonians, such as the second-quantized form of a vibrational Hamiltonian \cite{christiansen}.
This unified framework enables us to establish connections and to generalize existing decomposition methods.

To provide an appropriate context for comparison of different LCUs, let us consider the main contributions to their quantum resource cost. 
The LCU-based Hamiltonian oracle shown in Fig.~\ref{fig:lcu_oracle} comprises two primary components: the \prep circuit, which transforms the ancilla state $\ket{0}_a$ into a linear superposition with the LCU coefficients, and the \sel circuit, which applies the LCU unitaries in a controlled manner. In general, it is possible to move operations from \sel into \prep and vice-versa, and for a given LCU there might be several different associated oracles that realize the block-encoding. The unitaries in \sel correspond to multiplexed operators \cite{cnots,multiplex}, and their quantum gate cost depends on that of the associated unitaries $\hat U_k$ and a unary iteration over $k$ indices \cite{qrom}. The unary iteration is analogous to a for-loop over indices $k$ conditionally applying $\hat U_k$, and it can be efficiently incorporated by using a binary tree algorithm \cite{binary_tree_prepare}. This incurs an overhead scaling of $4(K-1)$ T-gates and $\myceil{\log_2(K)} - 1$ additional ancilla qubits. However, specific structures within the LCU decomposition can enable more efficient \sel circuits. We will explore this aspect further after introducing the different LCU decomposition techniques.

The rest of this paper is organized as follows. Section~\ref{sec:decomps} presents
 the unified \gls{MTD} framework for \gls{LCU} decompositions and a new LCU decomposition. 
 Then a list of recently proposed \gls{LCU} decompositions expressed as \glspl{MTD} is given in Sec.~\ref{sec:curr}. Finally, benchmarks for small systems and concluding remarks are offered in Sec.~\ref{sec:discussion} . 

\section{Majorana Tensor Decomposition Framework} \label{sec:decomps}

We will use the Majorana operators formalism because it provides a convenient bridge between 
fermionic and qubit operators. Majorana operators are defined as
\begin{align}
    \hat \gamma_{j\sigma,0} &\equiv \hat a_{j\sigma} + \hat a^\dagger_{j\sigma} \\
    \hat \gamma_{j\sigma,1} &\equiv i\left(\hat a^\dagger_{j\sigma} - \hat a_{j\sigma}\right),
\end{align}
with the algebraic relations
\begin{align}
    \{\hat \gamma_p, \hat \gamma_q\} &= 2\delta_{pq}\hat 1 \\
    \hat \gamma_p^\dagger &= \hat \gamma_p \\
    \hat \gamma_p^2 &= \hat 1,
\end{align}
where we used complex indices $p$ and $q$ containing sub-indices $\{j\sigma,m\}$: $m\in\{0,1\}$, $j$ is a spacial orbital index, and $\sigma\in\{\alpha,\beta\}$ is the spin-z projection. The electronic structure Hamiltonian for closed-shell systems can be written in the Majorana representation as \cite{majorana_l1,loaiza_lcu}
\begin{align}
    \hat H &= \left( \sum_i h_{ii} + \sum_{ij} g_{iijj}\right) \hat 1 \nonumber \\
    &+ \frac{i}{2}\sum_{\sigma}\sum_{ij} \left( h_{ij} + 2\sum_k g_{ijkk} \right) \hat \gamma_{i\sigma,0}\hat\gamma_{j\sigma,1} \nonumber \\
    &-\frac{1}{4}\sum_{\sigma\tau}\sum_{ijkl} g_{ijkl} \hat\gamma_{i\sigma,0}\hat\gamma_{j\sigma,1}\hat\gamma_{k\tau,0}\hat\gamma_{l\tau,1}. \label{eq:H_maj} \\
    &\equiv \hat H_0 + \hat H_1 + \hat H_2,
\end{align}
where we have respectively defined the zero-, two-, and four-Majorana components $\hat H_0$, $\hat H_1$, and $\hat H_2$. 

Since no terms with more than four Majorana operators appear in $\hat H$, a general \gls{MTD} representation of the Hamiltonian 
can be written as
\begin{equation} \label{eq:MTD}
    \hat H = \sum_{\substack{w_1,...,w_L \\ m_1,...,m_L}} \Omega_{\vec w, \vec m} \prod_{\nu=1}^{L} \hat V^{(\nu)\dagger}_{\vec m} \hat p_{\vec w}^{(\nu)} \hat V^{(\nu)}_{\vec m},
\end{equation}
where $\Omega_{\vec w, \vec m}$ is a tensor with possibly complex coefficients, $\hat p_{\vec w}^{(\nu)}$ is a unitary operator expressed as a polynomial of Majorana operators of degree $\leq 4$ with a minimal number of terms, $\hat V^{(\nu)}_{\vec m} \equiv e^{\sum_{p>q} \theta^{(\vec m,\nu)}_{pq} \hat \gamma_p \hat \gamma_q}\in \textrm{Spin}(4N)$ are unitary rotations that conserve the degree of $\hat p_{\vec w}^{(\nu)}$. We also constrain the product $\prod_{\nu=1}^{L}\hat p_{\vec w}^{(\nu)}$ to yield a Majorana polynomial of degree $\leq 4$ for all $w_1,...,w_{L}$ appearing in the sum, which restricts $L\leq 4$. For all decompositions shown in this work only spacial orbital rotations are used, which form a subgroup of $\textrm{Spin}(4N)$  
    \begin{align}
        \hat U_{\vec\theta} &\equiv \exp\left[\sum_{i>j} \theta_{ij} (\hat F^i_j - \hat F^j_i)\right] \in \rm{Spin}(N) \\
        &= \exp \left[\sum_{i>j} \frac{\theta_{ij}}{2} \sum_\sigma (\hat\gamma_{i\sigma,0}\hat\gamma_{j\sigma,0} + \hat\gamma_{i\sigma,1}\hat\gamma_{j\sigma,1})\right]. \label{eq:real_rot}
    \end{align}
 For example, in the linear case $\hat p_1^{\nu} = \gamma_{1\sigma,x}$, where $x=\{0,1\}$, all other linear in Majorana unitary operators can be obtained by conjugating with $\hat U_{\vec\theta}$ these two polynomials. For higher polynomial degrees, there is a still relatively low number of distinct $\hat p_{\vec w}^{(\nu)}$'s that give non-overlapping sets of unitary operators upon $\hat U_{\vec\theta}$ conjugation.  

Note that even though $\hat H$ could be in principle decomposed with unitaries that are Majorana polynomials of degree $>4$, these contributions would have to cancel to recover $\hat H$. Since there is an exponential number of Majorana polynomials with degree larger than $4$, keeping track of these operators and enforcing their cancellation is a computationally daunting task. By enforcing conservation of the Majorana polynomial degree and starting with polynomials of degree $\leq 4$, we do not have to consider an exponentially large space, making the space of unitaries for the \gls{LCU} more computationally tractable.


\subsection{MTD-$L^4$ decomposition}

Here we start with a novel LCU decomposition, \gls{MTD}-$L^4$, that has not been suggested before. The idea behind the \gls{MTD}-$L^4$ decomposition is representing unitaries as products of four unitaries that are first degree polynomials of Majoranas, thus having $L=4$. A way of writing this decomposition that conserves the spin-symmetry in Eq.~\eqref{eq:el_ham} is
\begin{equation} \label{eq:1pow4}
    \hat H_2 = \sum_{\vec m} \Omega_{\vec m} \sum_{\sigma\tau} \hat q_{\vec m\sigma, 0}^{(1)} \hat q_{\vec m\sigma, 1}^{(2)} \hat q_{\vec m\tau, 0}^{(3)} \hat q_{\vec m\tau, 1}^{(4)}, 
\end{equation}
where 
\begin{equation}
    \hat q_{\vec m\sigma, x}^{(\nu)} = \hat U^{(\nu)\dagger}_{\vec m}\hat \gamma_{1\sigma,x} \hat U^{(\nu)}_{\vec m}
\end{equation}
for $x=0,1$. Here, we skip the $\hat H_1$ part since there is a known 
optimal treatment of this part, and it will be discussed with other fermionic LCU decomposition techniques in Sec.~\ref{sec:ferm}. 
Note that real orbital rotations $\hat U$'s [Eq.~\eqref{eq:real_rot}] do not introduce Majoranas with different $x$ or $\sigma$ indices:
\begin{equation}
    \hat U^\dagger \hat\gamma_{i\sigma,x} \hat U = \sum_j U_{ij} \hat\gamma_{j\sigma,x}. \label{eq:maj_rotation}
\end{equation}
Thus, the \gls{MTD}-$L^4$ decomposition expresses the four-Majorana tensor, with associated rank = $4$, as a linear combination of outer products of four rank-1 vectors, with a trivial index $\vec w = 1$. Generally, the orbital rotations can be implemented as products of Givens rotations with $\mathcal{O}(N^2)$ operations for rotating $N$ orbitals. However, the orbital rotations shown here only act on a single spacial orbital, thus, only $\mathcal{O}(N)$ operations are necessary to implement these $\hat U$'s using Givens rotations. The corresponding angles of the Givens rotations can be found once the coefficients in Eq.~\eqref{eq:maj_rotation} have been specified; a detailed discussion of how to find these coefficients is shown in Sec.~\ref{subsec:AC}.

Next, we show different ways of obtaining the MTD-$L^4$ decomposition using common tensor decomposition techniques.

\subsubsection{MPS-based MTD-$L^4$} \label{subsec:mtd_svd}

\begin{figure*}[t]
    \centering
    \includegraphics[width=14cm]{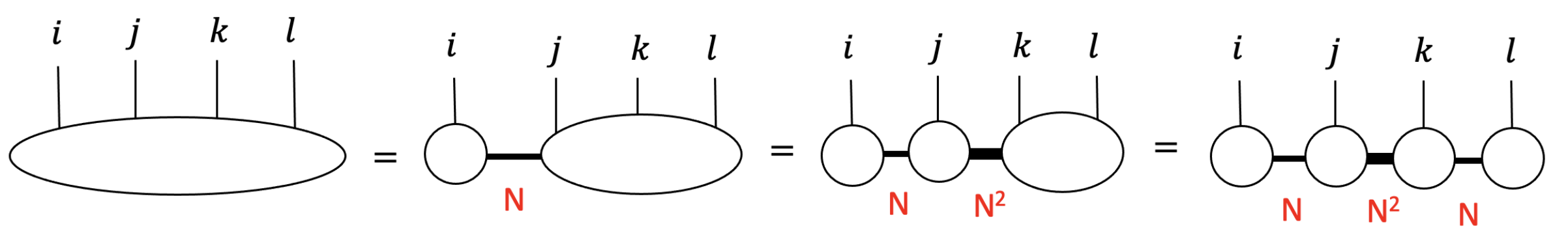}
    \caption{\gls{MPS} diagram representation of iterative factorization of two-electron tensor with nested \gls{SVD}s. Numbers in red show bond dimension for the matrices in the \gls{MPS}.}
    \label{fig:SVD_MPS}
\end{figure*}

\Gls{MPS} formalism \cite{dmrg_mps} applied to 4-index tensor $g_{pqrs}$ can be seen as iterative application of the \gls{SVD} (see Fig.(\ref{fig:SVD_MPS})). 
 The \gls{SVD} of a matrix is
\begin{equation}
    A_{uv} = \sum_{\alpha} U_{u\alpha} S_\alpha W_{v\alpha}.
\end{equation}
Treating the $g_{ijkl}$ tensor as a matrix with indices $i$ and $jkl$ allows to start the decomposition as follows
\begin{align}
    g_{ijkl} &= \sum_{u=1}^N U^{(1)}_{iu} S^{(1)}_{u} W^{(1)}_{jklu} \\
    &= \sum_{u=1}^N\sum_{v=1}^{N^2} S^{(1)}_{u}S^{(2)}_{v} U^{(1)}_{iu}U^{(2,u)}_{jv} W^{(2)}_{klv} \\
    &= \sum_{u,w=1}^N\sum_{v=1}^{N^2} S^{(1)}_{u}S^{(2)}_{v}S^{(3)}_{w} U^{(1)}_{iu}U^{(2,u)}_{jv}U^{(3,v)}_{kw} W^{(3)}_{lw}. \label{eq:mps}
\end{align}
From this, we can write
\begin{widetext}
\begin{align}
    \hat H_2 &= \sum_{ijkl}\sum_{uvw} \Omega_{uvw} U^{(1)}_{iu} U^{(2,u)}_{jv} U^{(3,v)}_{kw} W^{(3)}_{lw}\sum_{\sigma\tau}\hat\gamma_{i\sigma,0}\hat\gamma_{j\sigma,1}\hat\gamma_{k\tau,0}\hat\gamma_{l\tau,1} \\
    &= \sum_{uvw} \Omega_{uvw} \sum_{\sigma\tau} \hat q^{(1)}_{u,\sigma} \hat q^{(2)}_{uv,\sigma} \hat q^{(3)}_{vw,\tau} \hat q^{(4)}_{w,\tau},
\end{align}
\end{widetext}
where we use the notation for the rotated polynomials
\begin{align}
    \hat q^{(1)}_{u,\sigma} &\equiv \sum_i U^{(1)}_{iu} \hat\gamma_{i\sigma,0} \equiv \hat U_{1,u}^\dagger \hat\gamma_{1\sigma,0}\hat U_{1,u} \\
    \hat q^{(2)}_{uv,\sigma} &\equiv \sum_j U^{(2,u)}_{jv} \hat\gamma_{j\sigma,1} \equiv \hat U_{2,uv}^\dagger \hat\gamma_{1\sigma,1}\hat U_{2,uv}\\
    \hat q^{(3)}_{vw,\tau} &\equiv \sum_k U^{(3,v)}_{kw} \hat\gamma_{k\tau,0} \equiv \hat U_{3,vw}^\dagger \hat\gamma_{1\tau,0}\hat U_{3,vw}\\
    \hat q^{(4)}_{w,\tau} &\equiv \sum_l W^{(3)}_{lw} \hat\gamma_{l\tau,1} \equiv \hat U_{4,w}^\dagger \hat\gamma_{1\tau,1}\hat U_{4,w}
\end{align}
and $\Omega_{uvw} \equiv S^{(1)}_{u}S^{(2)}_{v}S^{(3)}_{w}$. Since in this case the \gls{SVD} yields real unitary matrices $U$ and $W$, it follows that for all of these polynomials $\hat q = \sum_p c_p \hat\gamma_p$, $c_p\in\mathbb{R}$ and $\sum_p |c_p|^2 = 1$, thus proving that these are real orbital rotations $\in \textrm{Spin}(N)$. The iterative \gls{SVD} procedure for the $g_{ijkl}$ tensor can be done efficiently on a classical computer, producing all necessary quantities for obtaining the \gls{LCU} and specifying the corresponding unitaries. We will refer to this decomposition as the MPS decomposition. Note that the MPS decomposition can be done also by first factorizing the middle index with $N^2$ components. This decomposition would be extremely similar to the \gls{DF} factorization of the two-electron tensor, although with a different LCU representation of the Hamiltonian.

\subsubsection{CP4-based \gls{MTD}-$L^4$}
\begin{figure}
    \centering
    $$\vcenter{\hbox{\includegraphics[width=3cm]{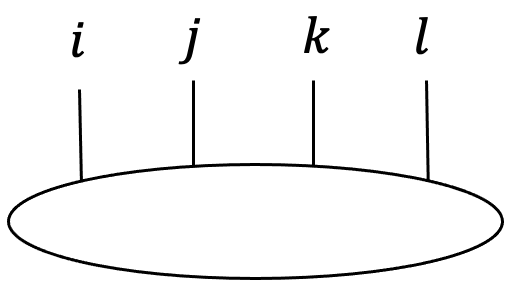}}} = \sum_m \Omega_m \left( \vcenter{\hbox{\includegraphics[width=3cm]{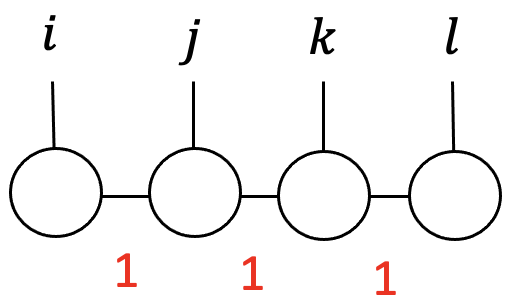}}}\right)^{(m)}$$
    \caption{MPS representation of CP4 factorization.}
    \label{fig:CP4_MPS}
\end{figure}

Alternatively the $g_{ijkl}$ tensor can be decomposed as a linear combination of outer products of rank-1 vectors based on the CP4 tensor decomposition\cite{cp_rev,cp,parafac,candecomp,polyadic,tcm}
\begin{equation}
    g_{ijkl} = \sum_m \Omega_m v^{(1,m)}_{i} v^{(2,m)}_{j} v^{(3,m)}_{k} v^{(4,m)}_{l},
\end{equation}
where each $\vec v^{(n,m)}$ for $n=1,2,3,4$ is an $N$-dimensional normalized vector. The associated \gls{MPS} representation would correspond to a linear combination of \glspl{MPS} with bond dimension 1, and is shown in Fig.\ref{fig:CP4_MPS}.
Note that the normalization can be done without loss of generality since all normalization constants can be absorbed in the $\Omega_m$ coefficient.  Obtaining the CP4 decomposition of the $g_{ijkl}$ tensor also defines an \gls{MTD}-$L^4$ decomposition as shown in Eq.~\eqref{eq:1pow4}, with the polynomials
\begin{align}
    \hat q_{m,\sigma}^{(1)} &\equiv \sum_i v_{i}^{(1,m)} \hat\gamma_{i\sigma,0} = \hat U^{(1)\dagger}_m \hat\gamma_{1\sigma,0} \hat U^{(1)}_m \\
    \hat q_{m,\sigma}^{(2)} &\equiv \sum_j v_{j}^{(2,m)} \hat\gamma_{j\sigma,1} = \hat U^{(2)\dagger}_m \hat\gamma_{1\sigma,1} \hat U^{(2)}_m \\
    \hat q_{m,\tau}^{(3)} &\equiv \sum_k v_{k}^{(3,m)} \hat\gamma_{k\tau,0} = \hat U^{(3)\dagger}_m \hat\gamma_{1\tau,0} \hat U^{(3)}_m \\
    \hat q_{m,\tau}^{(4)} &\equiv \sum_l v_{l}^{(4,m)} \hat\gamma_{l\tau,1} = \hat U^{(4)\dagger}_m \hat\gamma_{1\tau,1} \hat U^{(4)}_m,
\end{align}
which are implemented using ${\rm Spin}(N)$ rotations as shown for the MPS LCU. 

\subsubsection{SVD-based MTD-$L^4$}

There is yet another alternative scheme of iterative SVD for performing MTD-$L^4$.
This scheme groups indices differently than the MPS scheme, 
\begin{align}
    g_{ijkl} &= \sum_{\alpha_1=1}^N U_{i\alpha_1}^{(1)} S^{(1)}_{\alpha_1} V^{(1,\alpha_1)}_{jkl} \\
    &= \sum_{\alpha_1=1}^N U_{i\alpha_1}^{(1)} S^{(1)}_{\alpha_1} \sum_{\alpha_2=1}^N U_{j\alpha_2}^{(2,\alpha_1)} S^{(2,\alpha_1)}_{\alpha_2} V^{(2,\alpha_1\alpha_2)}_{kl} \\
    &= \sum_{\alpha_1=1}^N U_{i\alpha_1}^{(1)} S^{(1)}_{\alpha_1} \sum_{\alpha_2=1}^N U_{j\alpha_2}^{(2)} S^{(2,\alpha_1)}_{\alpha_2} \nonumber \\ &\ \ \times \sum_{\alpha_3=1}^N U_{k\alpha_3}^{(3)} S^{(3,\alpha_1\alpha_2)}_{\alpha_3} V^{(3,\alpha_1\alpha_2\alpha_3)}_{l}
\end{align}
This decomposition gives rise to an MTD for which $\vec m\cong(\alpha_1,\alpha_2,\alpha_3)$ with $N^3$ elements. This MTD is then determined by
\begin{align}
    \hat q^{(1)}_{\alpha_1,\sigma} &= \sum_i U_{i\alpha_1}^{(1)} \hat\gamma_{i\sigma,0} \\
    \hat q^{(2)}_{\alpha_1\alpha_2,\sigma} &= \sum_j U_{j\alpha_2}^{(2,\alpha_1)} \hat\gamma_{j\sigma,1} \\
    \hat q^{(3)}_{\alpha_1\alpha_2\alpha_3,\tau} &= \sum_k U_{k\alpha_3}^{(3,\alpha_1\alpha_2)} \hat\gamma_{k\tau,0} \\
    \hat q^{(4)}_{\alpha_1\alpha_2\alpha_3,\tau} &= \sum_l V_{l}^{(3,\alpha_1\alpha_2\alpha_3)} \hat\gamma_{l\tau,1} \\
    \Omega_{\alpha_1\alpha_2\alpha_3} &= S^{(1)}_{\alpha_1} S^{(2,\alpha_1)}_{\alpha_2} S^{(3,\alpha_1\alpha_2)}_{\alpha_3}.
\end{align}
Note that if we ``flatten" the $N\times N \times N$ vector $(\alpha_1,\alpha_2,\alpha_3)$ into a one-dimensional array with $N^3$ elements, we recover an MTD which is identical to the MTD-CP4, the only difference being how the tensor was decomposed. We refer to this family of decompositions as MTD-$L^4$. However, the additional structure in some of these decompositions (e.g., MPS) make it amenable to more efficient LCU oracles, as shown in the circuits in the Appendix A.

\section{Current LCU\lowercase{s} as MTD\lowercase{s}} \label{sec:curr}
Here we show how current \glspl{LCU} can be written as \glspl{MTD}, these mappings are summarized in Table~\ref{tab:curr_4}. 

\subsection{Qubit-based decompositions}
\subsubsection{Pauli product decompositions}
Also known as the sparse \gls{LCU}, the simplest \gls{LCU} decomposition can be obtained by mapping the electronic structure Hamiltonian [Eq.~\eqref{eq:el_ham}] into qubit operators through the use of some fermion-to-qubit mapping (e.g., Jordan-Wigner \cite{jordan_wigner} or Bravyi-Kitaev \cite{bk1,bk2,bk3}):
\begin{equation} \label{eq:q_pauli}
    \hat H = \sum_q d_q \hat P_q,
\end{equation}
where $d_q$ are coefficients for the electronic structure Hamiltonian, and $\hat P_q$ are products of Pauli operators over qubits, commonly referred to as Pauli products. 

To see the correspondence between electron integrals in the fermionic Hamiltonian Eq.~ \eqref{eq:el_ham} and coefficients of Pauli products ($d_q$), it is convenient to write the fermionic Hamiltonian using the reflection operators $\hat Q_{kl\sigma} \equiv i \hat\gamma_{k\sigma,0}\hat\gamma_{l\sigma,1}$:
\begin{equation} \label{eq:q_ham}
    \hat H = \hat H_0 + \frac{1}{2}\sum_{ij} \tilde h_{ij} \sum_\sigma \hat Q_{ij\sigma} + \frac{1}{4}\sum_{ijkl}g_{ijkl} \sum_{\sigma\tau} \hat Q_{ij\sigma} \hat Q_{kl\tau},
\end{equation}
where we have defined $\tilde h_{ij} \equiv h_{ij} + 2\sum_k g_{ijkk}$. This expression is an LCU corresponding to the sparse LCU with $\tilde h_{ij}/2$ and $g_{ijkl}/4$ equal to 
$d_q$ in \eqref{eq:q_pauli}. 

The associated \gls{MTD} representation [Eq.~\eqref{eq:MTD}] has $L=1$, and can be obtained by a trivial $\vec m=1$ index with a single element and no rotation, i.e. $\hat V_{1} = \hat 1$. There will be a single $\vec w\rightarrow w$ coefficient, thus making the coefficients tensor $\Omega_{w}$,
with the corresponding $\Omega_{w}$/$\hat p_{w}$'s running over each of the coefficients/operators appearing in Eq.~\eqref{eq:q_ham}.

A subtle point is that the two-electron component of $\hat H_e$ can be expressed by separating $\sigma=\tau$ and $\sigma\neq\tau$ components
\begin{align}
    \hat H_2 &= \left(\frac{1}{2}\sum_{ij}g_{ijji}\right)\hat 1 \nonumber \\
    &\ +\frac{1}{4}\sum_{\sigma\neq\tau}\sum_{ijkl}g_{ijkl}\hat Q_{ij\sigma} \hat Q_{kl\tau} \nonumber \\
    &\ + \frac{1}{2}\sum_\sigma\sum_{i>k,l>j}(g_{ijkl}-g_{ilkj})\hat Q_{ij\sigma}\hat Q_{kl\sigma}.
\end{align}
This form of $\hat H_2$ gives lower 1-norms than that in Eq.~\eqref{eq:q_ham}\cite{majorana_l1}, 
but it requires separating 
the spin components and as a consequence twice as many coefficients to be loaded by the \prep circuit, making it generally more expensive. Thus, we will always work with the Hamiltonian form shown in Eq.~\eqref{eq:H_maj}, making the two-electron tensor representable using spacial orbitals. In the case where this spin-symmetry is broken, as is the case when obtaining a Hamiltonian from unrestricted Hartree-Fock calculations and generally for any open-shell system, only non-zero coefficients of the two-electron tensor in spin-orbitals need to be loaded. This can be done by using a contiguous register, as shown in Appendix A for the sparse LCU.

\subsubsection{Anti-commuting groupings} \label{subsec:AC}
The 1-norm of the Pauli decomposition can be reduced by grouping Pauli products into larger unitaries. Normalized linear combinations of mutually anti-commuting Pauli products yield unitary operators \cite{anticommuting} that improve on the 1-norm of \eqref{eq:q_pauli} 
\begin{equation} \label{eq:H_ac}
    \hat H = \sum_{n=1}^{N_{\rm{AC}}} a_n \hat A_n,
\end{equation}
where $N_{\rm{AC}}$ represents the total number of groups, and
\begin{align}
a_n &\equiv \sqrt{\sum_{q\in K_n} |d_q|^2}, \\
\hat A_n &\equiv \frac{1}{a_n} \sum_{q\in K_n} d_q \hat P_q,
\end{align}
with $\hat A_n$ representing linear combinations of mutually anti-commuting Pauli products, and $K_n$ denoting the associated indices. Within the \gls{MTD} framework [Eq.~\eqref{eq:MTD}], this decomposition has $L=1$, where $\vec m$ takes on a single element such that $\hat V_{1} = \hat 1$. The $\vec w$ index is identical to the $n$ index in Eq.~\eqref{eq:H_ac}, with $\Omega_{n} = a_{n}$ and $\hat p_{n} = \hat A_n$. Referring to Eq.~\eqref{eq:q_ham}, we observe that the coefficients $d_q$ associated with products of two Majoranas are purely imaginary, whereas those arising from four Majoranas are real. Considering that products of two Majoranas are anti-Hermitian and products of four Majoranas are Hermitian, it follows that $\hat A_n^\dagger = \hat A_n$, indicating that $\hat A_n$ operators are reflection operators. Lastly, each $\hat A_n$ can be implemented using the following equality:
\begin{equation} \label{eq:naive_ac}
\hat A_n = \prod_{q\in K_n\uparrow} e^{i\phi_q^{(n)} \hat P_q} \prod_{q\in K_n\downarrow} e^{i\phi_q^{(n)} \hat P_q},
\end{equation}
where $\uparrow (\downarrow)$ denotes ascending (descending) order through $K_n$, and
\begin{equation}
\phi_q^{(n)} \equiv \frac{1}{2}\arcsin\frac{d_q}{\sqrt{\sum\limits_{\substack{i\in K_n \\ i\leq q}} d_i^2}}.
\end{equation}
Note that this decomposition is the only one presented here in which unitaries can combine both two-Majorana and four-Majorana terms. The application of controlled $\hat A_n$ unitaries, as shown in Eq.~\eqref{eq:naive_ac}, would require $2|K_n|$ controlled exponentials of Pauli operators, and as a result high T-gate cost. To bring this cost down, we introduce an alternative approach that reduces this number to only
 one controlled unitary. The key idea is that for each group $K_n$, the anti-commutativity of the $|K_n|$ Pauli products generates the Clifford algebra isomorphic to that of the same number of Majorana operators. The key property of a normalized linear combination of $N$ single Majorana operators is that it can be written as a single Majorana conjugated by a sequence of Givens orbital rotations:
\begin{equation}
    \sum_n c_n \hat \gamma_n = \left( \prod_{n=N-1}^{1}e^{\theta_n \hat\gamma_n \hat\gamma_{n+1}} \right) \hat\gamma_1 \left( \prod_{n=1}^{N-1}e^{-\theta_n \hat\gamma_n \hat\gamma_{n+1}} \right),
\end{equation}
where the angles $\theta_n$ will depend on the $c_n$'s as 
\begin{align}
    \theta_1 &= \frac{1}{2}\arccos c_1, \nonumber \\
    \theta_2 &= \frac{1}{2} \arccos \left(\frac{c_2}{\sin 2\theta_1} \right), \nonumber \\
    \theta_3 &= \frac{1}{2} \arccos \left(\frac{c_3}{\prod_{j \leq 2} \sin 2\theta_j} \right), \nonumber \\
    &\hspace{2cm} \vdots \nonumber \\
    \theta_{N-2} &= \frac{1}{2} \arccos \left(\frac{c_{N-2}}{\prod_{j \leq (N-3)} \sin 2\theta_j} \right), \nonumber \\
    \theta_{N-1} &= \frac{{\rm sign}(c_{N})}{2} \arccos \left(\frac{c_{N-1}}{\prod_{j \leq N-2} \sin 2\theta_j} \right).
\end{align}
Given the identical algebraic structure, the same Givens-based diagonalization can be used to express any unitary consisting of mutually anti-commuting Paulis $\hat A \equiv \sum_{j=1}^J a_j \hat P_j$:
\begin{equation}
    \hat A = \left( \prod_{j=J - 1}^1 e^{\theta_j \hat P_j \hat P_{j+1}} \right) \hat P_1 \left( \prod_{j=1}^{J-1} e^{-\theta_j \hat P_j \hat P_{j+1}} \right).
\end{equation}
By using this expression, the controlled application of each $\hat A_n$  only requires a single controlled operation, as shown in the \sel circuit for this LCU in Appendix A.

\subsubsection{Orbital-optimized qubit operators}
1-norm and the number of unitaries for qubit-based decompositions can be reduced further by working with an orbital basis that differs from the canonical molecular orbitals obtained from solving the Hartree-Fock equations. In this study, we explored two distinct orbital rotation schemes: the Foster-Boys orbitals \cite{FB} and optimized orbitals obtained by directly minimizing the 1-norm of the LCU. In both cases, the Hamiltonian is transformed as $\hat U^\dagger \hat H \hat U$, where $\hat U$ represents the chosen orbital rotation. This amounts to choosing the Majorana-conserving rotation $\hat V=\hat U$ instead to $\hat V=1$ in the \gls{MTD} expression [Eq.~\eqref{eq:MTD}]. For the systems studied in Ref.~\citenum{majorana_l1}, the 1-norm scaling for the Foster-Boys orbitals closely approaches that of the optimal rotations, while requiring $\mathcal{O}(N^3)$ operations to find $\hat U$, as opposed to the $\mathcal{O}(N^5)$ operations needed for finding the 1-norm minimizing rotation. 

\subsection{Fermionic LCU decompositions}\label{sec:ferm}
Next, we illustrate how fermionic-based LCU decompositions can be expressed as \gls{MTD}. Unitaries in these \glspl{LCU} are usually constructed as orbital rotations acting on reflections $\hat Q_{ii\sigma}\equiv i \hat\gamma_{i\sigma,0}\hat\gamma_{i\sigma,1} =2\hat{n}_{i \sigma}-\hat{1}$, 
where $\hat n_{i\sigma}\equiv \hat a_{i\sigma}^\dagger\hat a_{i\sigma}$ represents the occupation number operator. The fermion-to-qubit mappings transform $Q_{ii\sigma}$ operator to the qubit operator $\hat z_{i\sigma}$.

Some methods, particularly those that factorize the $g_{ijkl}$ tensor, separate the one-electron and two-electron operators, which amounts to having separate decompositions for the two-Majorana $\left(\hat H_1\right)$ and four-Majorana $\left(\hat H_2\right)$ parts in Eq.~\eqref{eq:H_maj}, with the decomposition for $\hat H_1$ being capped at $L=2$ and polynomials $\hat p$'s with degree $\leq 2$. Since one-electron operators can be efficiently diagonalized and implemented with an optimal 1-norm \cite{loaiza_lcu}, any decomposition that treats $\hat H_1$ and $\hat H_2$ separately has an optimal encoding of $\hat H_1$ as
\begin{align} \label{eq:1el_diag}
    \hat H_1 &= \hat U^\dagger \left(\sum_i \mu_i\sum_\sigma \hat Q_{ii\sigma} \right) \hat U.
\end{align}

Note that the most efficient encodings for many fermionic-based decompositions, such as \gls{DF} and \gls{THC}, merge the one- and two-electron coefficients, using a quantum register to indicate whether a given term is coming from one- or two-electron operators \cite{THC}. 
In the case of $\hat H_2$, the coefficient invariance in the Hamiltonian with respect to spin enables us to decompose the two-electron tensor $g_{ijkl}$ using the spacial-orbital indices, effectively working in the algebra defined by the $\hat F^i_j$ operators. Apart from computational efficiency, operating within this spin-symmetric algebra facilitates compilation of the LCU unitaries while requiring less data to be loaded on the quantum computer, as demonstrated in Ref.~\citenum{THC}. 

\subsubsection{Single factorization}
Although originally deduced with fermionic operators \cite{sf}, the \gls{SF} technique can be written with Majoranas 
as \cite{majorana_l1,THC}:
\begin{align}
    \hat H &= c_{\rm{SF}}\hat 1 + \frac{i}{2}\sum_\sigma \sum_{ij}\tilde h_{ij} \hat\gamma_{i\sigma,0}\hat \gamma_{j\sigma,1} \nonumber \\
    &\ \ - \frac{1}{4}\sum_{\mathcal{l}=1}^{N^2} \left(\sum_\sigma \sum_{ij} W_{ij}^{(\mathcal{l})} \hat\gamma_{i\sigma,0}\hat\gamma_{j\sigma,1} \right)^2, \label{eq:H_sf}
\end{align}
where $W_{ij}^{(\mathcal{l})}$ represents the Cholesky decomposition of the $g_{ijkl}$ tensor written as a $N^2\times N^2$ matrix $g_{ijkl}$, and $c_{\rm{SF}}$ is a constant. We note that the $\mathcal{l}$ index is usually truncated at some $M \ll N^2$.  Equation \eqref{eq:H_sf} illustrates how this decomposition can be cast as an \gls{MTD} that separates $\hat H_1$ and $\hat H_2$, with $L=2$ [Eq.~\eqref{eq:MTD}]. The associated sum over $\vec m$ has trivial rotations  $\hat V = \hat 1$ and $M$ elements. The Majorana polynomials $\hat p$'s correspond to the Majorana products in Eq.~\eqref{eq:H_sf}. Additionally, the complete-square structure of the two-electron elements allows them to be expressed as squares of $\hat p_{\mathcal{l}} \equiv \sum_\sigma\sum_{ij} W_{ij}^{(\mathcal{l})} \hat\gamma_{i\sigma,0}\hat\gamma_{j\sigma,1}$ operators. Complete-squares can be encoded through the use of oblivious amplitude amplification, which implements through qubitization the second-degree Chebyshev polynomial of the normalized $\hat p_{\mathcal{l}}$, i.e. $2\left(\hat p_{\mathcal{l}}/N_{\mathcal{l}}^{(\rm{SF})}\right)^2 - \hat 1$, where the normalization constant $N_{\mathcal{l}}^{(\rm{SF})} \equiv 2\sum_{ij} |W_{ij}^{(\mathcal{l})}|$. Finally, we note that the original procedure did not diagonalize the one-electron term. The \gls{MTD} representation of single factorization becomes
\begin{align} 
    \hat H &= \tilde c_{\rm{SF}} \hat 1 + i\sum_{jk\sigma} \Lambda_{jk} \hat \gamma_{j\sigma,0} \hat\gamma_{k\sigma,1} \nonumber \\
    &\ \ + \sum_{\mathcal{l}}\Omega_{\mathcal{l}} \left(2\left(\frac{\hat p_{\mathcal{l}}}{N_{\mathcal{l}}^{(\rm{SF})}}\right)^2 - \hat 1\right), \label{eq:sf_mtd}
\end{align}
with $\Lambda_{jk} \equiv \frac{h_{jk}}{2} + \sum_{l} g_{jkll}$, $\Omega_{\mathcal{l}} \equiv -\frac{\left(N_{\mathcal{l}}^{(\rm{SF})}\right)^2}{8}$, and $\tilde c_{\rm{SF}}$ has been adjusted to account for the identity term in the Chebyshev polynomials. 

\subsubsection{Double factorization}
The \gls{DF} decomposition \cite{df_1,df_2,df_3,df_4,df_5,femoco_df,qubitization_df} can be viewed as an extension of \gls{SF} with the distinction that two-Majorana polynomials are first diagonalized to achieve an optimal 1-norm. In the \gls{DF} approach, the one-electron term is diagonalized as shown in Eq.~\eqref{eq:1el_diag}. This diagonalization is also used to implement the two-Majorana polynomials $\hat p_{\mathcal{l}}$. This approach has $L=1$ for the \gls{MTD} [Eq.~\eqref{eq:MTD}]. Defining the diagonalized unitaries
\begin{align}
    \hat p_{\mathcal{l}} &\equiv i\sum_{ij} W^{(\mathcal{l})}_{ij} \sum_\sigma \hat\gamma_{i\sigma,0}\hat\gamma_{j\sigma,1} \nonumber \\
    &= i\hat U_{\mathcal{l}}^\dagger \left(\sum_i \mu^{(\mathcal{l})}_i \sum_\sigma \hat\gamma_{i\sigma,0}\hat\gamma_{i\sigma,1} \right) \hat U_{\mathcal{l}}
\end{align}
the \gls{DF} \gls{LCU} is written in the fermionic representation as
\begin{align}
    \hat H &= c_{\rm{DF}}\hat 1 + \sum_i \mu_i^{(0)}\sum_\sigma \hat U_0^\dagger \hat Q_{ii\sigma} \hat U_0 \nonumber \\
    &\ \ + \sum_{\mathcal{l}} \left (\sum_i \mu_i^{(\mathcal{l})} \sum_\sigma \hat U_{\mathcal{l}}^\dagger \hat Q_{ii\sigma} \hat U_{\mathcal{l}} \right )^2,
\end{align}
where $c_{\rm{DF}}$ is a constant that depends on the $\mu$ factors (see Ref.~\citenum{THC}). The $\mu^{(l)}$ vectors are obtained by diagonalizing the $W^{(\mathcal{l})}$ matrices appearing in the \gls{SF} approach.
This can be recast in the \gls{MTD} form:
\begin{align}
    \hat H &= c_{\rm{DF}}\hat 1 + i\sum_j \mu_j^{(0)} \sum_\sigma \hat U_0^\dagger \hat\gamma_{j\sigma,0}\hat\gamma_{j\sigma,1} \hat U_0 \nonumber \\
    &\ \ + \sum_{\mathcal{l}} \Omega_{\mathcal{l}} \hat U_{\mathcal{l}}^\dagger \left( 2\left(\frac{\hat p_{\mathcal{l}}}{N_{\mathcal{l}}^{(\rm{DF})}}\right)^2 - \hat 1 \right) \hat U_{\mathcal{l}},
\end{align}
with $N_{\mathcal{l}}^{(\rm{DF})} \equiv \sum_i |\mu_i^{(\mathcal{l})}| $, $\Omega_{\mathcal{l}} \equiv \frac{\left(N_{\mathcal{l}}^{(\rm{DF})}\right)^2}{2}$. For the \gls{MTD}, $\vec m$ runs over $M \leq N^2$ indices. 
However, $\vec m$ is equal to polynomial subscript $\mathcal{l}$ ($\hat p_{\mathcal{l}}$), which would correspond to an implicit Kronecker delta between $\vec m$ and $\vec w$ indices (here written as $\mathcal{l}$) in the $\Omega$ tensor, with $\hat V_{\mathcal{l}} = \hat U_{\mathcal{l}}\in \textrm{Spin}(N)$. 

\subsubsection{Cartan sub-algebra}
The Cartan sub-algebra (CSA) decomposition generalizes \gls{DF} to include two-electron terms that are not parts of a complete-square. Thus, it corresponds to a full-rank factorization, as opposed to \gls{DF} which is known as a low-rank factorization \cite{CSA,csa_constrained,parrish_csa}. CSA \gls{LCU} is
\begin{align}
    \hat H &= c_{\rm{CSA}}\hat 1 + \sum_i \mu_i\sum_\sigma \hat U_0^\dagger \hat Q_{ii\sigma} \hat U_0 \nonumber \\
    &\ \ + \sum_{\mathcal{l}} \sum_{ij} \lambda_{ij}^{(\mathcal{l})} \sum_{\sigma\tau}\hat U_{\mathcal{l}}^\dagger \hat Q_{ii\sigma}\hat Q_{jj\tau} \hat U_{\mathcal{l}},
\end{align}
where the one-electron term is the same as that appearing in the \gls{DF} framework. The corresponding \gls{MTD} form is
\begin{align}
    \hat H &= c_{\rm{CSA}}\hat 1 + i\sum_j \mu_j^{(0)} \sum_\sigma \hat U_0^\dagger \hat\gamma_{j\sigma,0}\hat\gamma_{j\sigma,1} \hat U_0 \nonumber \\
    &\ \ - \sum_{\mathcal{l},ij} \lambda_{ij}^{(\mathcal{l})} \sum_{\sigma\tau}\hat U_{\mathcal{l}}^\dagger \hat\gamma_{i\sigma,0}\hat\gamma_{i\sigma,1}\hat\gamma_{j\tau,0}\hat\gamma_{j\tau,1} \hat U_{\mathcal{l}},
\end{align}
with $L=2$ [Eq.~\eqref{eq:MTD}]. The associated $\Omega$ tensor has an implicit Kronecker delta in the $m_1$ and $m_2$ indices. Despite being more flexible than \gls{DF}, the non-linear optimization procedure required to obtain the Cartan sub-algebra decomposition becomes computationally expensive for large systems.

\subsubsection{Tensor Hypercontraction}
The \gls{THC} decomposition \cite{THC,thc_scaling} expresses the Hamiltonian as
\begin{align}
    \hat H &= c_{\rm{THC}}\hat 1 + i\sum_j \mu_j \sum_\sigma \hat U_0^\dagger \hat Q_{jj\sigma} \hat U_0 \nonumber \\
    &\ \ + \sum_{\mu,\nu=1}^{K} \zeta_{\mu\nu} \sum_\sigma\hat U_\mu^\dagger \hat Q_{11\sigma} \hat U_\mu \sum_\tau\hat U_\nu^\dagger \hat Q_{11\tau}\hat U_\nu, \label{thc}
\end{align}
which corresponds to an $L=2$ \gls{MTD} with 
$\vec w = 1$ index [Eq.~\eqref{eq:MTD}], the limit $K$ is referred as the rank of the THC decomposition. 
Since the $\hat U$'s orbital rotations are only acting on orbital $1$, their implementation requires $\mathcal{O}(N)$ Givens rotations, as opposed to the $\mathcal{O}(N^2)$ required for rotating all orbitals \cite{femoco_df}. To better understand this reduction in the number of Givens rotations, we start by noting that for an arbitrary rotation $U\in\textrm{Spin}(N)$, the maximal torus theorem \cite{Hall:MTT} allows us to write
\begin{equation}
    \hat U = \prod_{i>j=1}^N e^{\theta_{ij} \hat F^i_j},
\end{equation}
where the product over Givens rotations can be taken in any order, noting that different sets of $\theta_{ij}$'s will need to be considered for each different ordering. Since $\hat Q_{11\sigma}$ only has a component in orbital $1$ we can simplify its transformation by leaving only $\hat F^i_j$ generators where $i$ or $j$ is 1. To demonstrate validity of this simplification, we use a freedom to choose an order of exponential functions in the $\hat U$ product. Such a choice can always 
be made because generators $\hat F^i_j$ form a Lie algebra. The convenient order is to group all Givens rotations not acting on orbital $1$ on the left-hand side of $\hat U$, 
\bea\notag
    \hat U^\dagger \hat Q_{11\sigma} \hat U &=& e^{\theta_{12}\hat F^1_2}e^{\theta_{13}\hat F^1_3}...e^{\theta_{1N}\hat F^1_N} \left(\prod_{i>j>1} e^{\theta_{ij}\hat F^i_j} \right)^\dagger \hat Q_{11\sigma}\\ \notag
    &\times &\left(\prod_{i>j>1} e^{\theta_{ij}\hat F^i_j} \right) e^{\theta_{1N}\hat F^1_N} ... e^{\theta_{12}\hat F^1_2} \\ \notag
    &=& e^{\theta_{12}\hat F^1_2}e^{\theta_{13}\hat F^1_3}...e^{\theta_{1N}\hat F^1_N} \hat Q_{11\sigma} \\
    &\times & e^{\theta_{1N}\hat F^1_N} ... e^{\theta_{12}\hat F^1_2},
\eea
where any Givens rotation not involving orbital $1$ commutes with $\hat Q_{11\sigma}$. Evidently, $N-1$ Givens rotations are enough for arbitrary orbital rotation of $\hat Q_{11\sigma}$. \\


\begin{widetext}
\begin{sidewaystable}
\small
    \centering
    \label{tab:curr_4}
        \caption{\gls{MTD} representation of current \glspl{LCU}. Pauli and anticommuting (AC) groupings, with their orbital optimized (OO-) versions, correspond to decomposing the full Hamiltonian $\hat H$. All other decompositions separate the one-electron and two-electron components $\hat H_1$ and $\hat H_2$, with the $\hat H_2$ components shown here and $\hat H_1$ in Eq.\eqref{eq:1el_diag}. Unitaries marked with $\hat U$ correspond to real orbital rotations $\in \textrm{Spin}(N)$. All necessary definitions for operators and constants can be found in the section for the corresponding decomposition.}
    \begin{adjustbox}{scale=0.95,center}
    \begin{tabular}{|c|c|c|c|p{6cm}|p{4.5cm}|p{5.5cm}|} \hline
    LCU & $L$ & $\vec m$ &  $\hat V_{\vec m}$ & \hspace{3.2cm}$\vec w$ & \hspace{2.3cm}$\Omega_{\vec w}^{\vec m}$ & \hspace{2.7cm}$\hat p_{w_k}$  \\ \hline
    Pauli & 1 & $\vec m = 1$ & $\hat V_1 = \hat 1$ &\parbox{6.5cm}{\begin{equation} \nonumber
  \vec w =
    \begin{cases}
      (i,j,\sigma) & \rightarrow 2N^2\\
      (i,j,k,l,\sigma) & \rightarrow 2N^4 \\
      (i>k,l>j,\sigma) & \rightarrow N^2(N-1)^2 \\
    \end{cases}       
\end{equation} \\ $|\vec w| = 2N^2+2N^4+N^2(N-1)^2$} & \parbox{5cm}{\begin{equation} \nonumber
\begin{cases}
      \Omega_{ij\sigma} = \frac{h_{ij}}{2} + \sum_k g_{ijkk}\\
      \Omega_{ijkl\sigma} = -\frac{g_{ijkl}}{4}  \\
      \Omega_{ijkl\sigma} = \frac{g_{ilkj}-g_{ijkl}}{2}  \\
    \end{cases}       
\end{equation}} & \parbox{6cm}{\begin{equation} \nonumber
\begin{cases}
      \hat p_{jk\sigma} = i\hat\gamma_{j\sigma,0}\hat\gamma_{j\sigma,1}\\
      \hat p_{ijkl\sigma} = \hat\gamma_{i\sigma,0}\hat\gamma_{j\sigma,1}\hat\gamma_{k\overline\sigma,0}\hat\gamma_{l\overline\sigma,1}  \\
      \hat p_{ijkl\sigma} = \hat\gamma_{i\sigma,0}\hat\gamma_{j\sigma,1}\hat\gamma_{k\sigma,0}\hat\gamma_{l\sigma,1}  \\
    \end{cases}       
\end{equation}} \\ \hline
    AC & 1 & $\vec m = 1$ & $\hat V_{1} = \hat 1$ & $\vec w \equiv n = 1,...,N_{\rm{AC}}$ & $\Omega_{n} = a_n \equiv \sqrt{\sum_{q\in K_n} |d_q|^2}$ & $\hat p_n = \hat A_n \equiv \frac{1}{a_n}\sum_{q\in K_n} d_q \hat P_q$ \\ \hline
     OO-Pauli & 1 & $\vec m = 1$ & $\hat V_1 = \hat U_{\vec \theta}$ & Same as in Pauli & Same as in Pauli & Same as in Pauli \\ \hline
    OO-AC & 1 & $\vec m = 1$ & $\hat V_{1} = \hat U_{\vec \theta}$ & Same as in AC & Same as in AC & Same as in AC \\ \hline
    SF & 1 & $\vec m = 1$ & $\hat V_1=\hat 1$ & $\vec w=\mathcal{l}=1,...,M_{\rm{SF}}\leq N^2$ & \parbox{5cm}{$\Omega_{\mathcal{l}} = -\frac{N_{\mathcal{l}}^{(\rm{SF})}}{8}$\\$N_{\mathcal{l}}^{(\rm{SF})} = 2\sum_{ij}|W^{(\mathcal{l})}_{ij}|$\\ $g_{ijkl} = \sum_{\mathcal{l}} W_{ij}^{(\mathcal{l})} W_{kl}^{(\mathcal{l})}$} & \parbox{5.5cm}{$\hat p_{\mathcal{l}}+\hat 1 = \left(\frac{\sqrt{2}}{N_{\mathcal{l}}^{(\rm{SF})}}\sum_{ij\sigma} W_{ij}^{(\mathcal{l})} \hat\gamma_{i\sigma,0}\hat\gamma_{j\sigma,1}\right)^2$} \\ \hline
    DF & 1 & $\vec m = \mathcal{l} = 1,...,M_{\rm{DF}} \leq N^2$ & $\hat V_{\mathcal{l}}=\hat U_{\vec \theta_{\mathcal{l}}} \equiv \hat U_{\mathcal{l}}$ & $\vec w=\mathcal{l}=1,...,M_{\rm{DF}}\leq N^2$ & \parbox{5cm}{$\Omega_{\mathcal{l}} = -\frac{N_{\mathcal{l}}^{(\rm{DF})}}{2}$\\$N_{\mathcal{l}}^{(\rm{DF})} = \sum_{i}|\mu^{(\mathcal{l})}_{i}|$\\ $W_{ij}^{(\mathcal{l})} = \sum_{k} \mu^{(\mathcal{l})}_k U_{\mathcal{l},ik}U_{\mathcal{l},jk}$} & \parbox{5.5cm}{$\hat p_{\mathcal{l}}+\hat 1=  \left(\frac{\sqrt{2}}{N_{\mathcal{l}}^{(\rm{DF})}}\sum_{i\sigma} \mu_{i}^{(\mathcal{l})}\hat\gamma_{i\sigma,0}\hat\gamma_{i\sigma,1} \right)^2$}  \\ \hline
    CSA & 2 & $\vec m = \mathcal{l} = 1,...,M_{\rm{CSA}}$ & $\hat V_{\mathcal{l}}=\hat U_{\vec \theta_{\mathcal{l}}} \equiv \hat U_{\mathcal{l}}$ & $\vec w=(i,j,\sigma,\tau), i\sigma\neq j\tau \rightarrow (4N^2 - 2N)$ & \parbox{5cm}{$\Omega^{(\mathcal{l})}_{i\sigma j\tau} = \lambda^{(\mathcal{l})}_{ij}$} & \parbox{6cm}{$\hat p^{(1)}_{i\sigma} = \hat\gamma_{i\sigma,0}\hat\gamma_{i\sigma,1}$ \\ $\hat p^{(2)}_{j\tau} = \hat\gamma_{j\tau,0}\hat\gamma_{j\tau,1}$}  \\ \hline
    THC & 2 & \parbox{3cm}{$\vec m = (\mu,\nu) \rightarrow M_{\rm{THC}}^2$ \\ $\mu,\nu = 1,...,M_{\rm{THC}}$} & $\hat V_{\mu}=\hat U_{\mu}$ & $\vec w=(\sigma,\tau) \rightarrow (4)$ & \parbox{5cm}{$\Omega^{\mu\nu}_{\sigma\tau} = \zeta_{\mu\nu}$} & \parbox{6cm}{$\hat p^{(1)}_{\sigma} =\hat\gamma_{1\sigma,0}\hat\gamma_{1\sigma,1}$\\$\hat p^{(2)}_{\tau} =\hat\gamma_{1\tau,0}\hat\gamma_{1\tau,1}$}  \\ \hline
    $L^4$ & 4 & \parbox{3cm}{$\vec m = m = 1,\cdots,W$}  & $\hat V_m^{(\nu)} = \hat U^{(\nu)}_m$ & $\vec w=(\sigma,\tau)\rightarrow (4)$ & \parbox{5cm}{$\Omega_{\sigma\tau}^{(m)} = \Omega_m$} & \parbox{6cm}{$\hat p_{\sigma}^{(1)} = \hat\gamma_{1\sigma,0}$ \\ $\hat p_{\sigma}^{(2)} = \hat\gamma_{1\sigma,1}$ \\ $\hat p_{\tau}^{(3)} = \hat\gamma_{1\tau,0}$ \\ $\hat p_{\tau}^{(4)} = \hat\gamma_{1\tau,1}$} \\ \hline
    \end{tabular}

    \end{adjustbox}
\end{sidewaystable}
\end{widetext}

\section{Results and Discussion} \label{sec:discussion}

The most time-consuming part of the fault-tolerant QPE algorithm is performing T-gates and their error correction. Considering that the number of applications of the Hamiltonian LCU 
oracle is proportional to the 1-norm of the used LCU, as a quantum resource cost 
we use a product of the LCU 1-norm by the number of T-gates required for an implementation of 
the Hamiltonian oracle. 
The latter corresponds to the number of T-gates in one \sel and two \prep operations. 
We will refer to the quantum resource cost simply as hardness.  Additionally, we note that optimizing the compilation strategy for the $\mathtt{SELECT}$ circuit is imperative for minimizing the hardness. Compilation of these circuits typically exploit structures present in the used \gls{LCU} \textit{ansatz}, as seen in Ref.~\citenum{THC} for \gls{DF} and \gls{THC} approaches which make use of the spin-symmetry that is present (i.e. working with $\hat F^i_j$ operators). Whether better circuits can be found for the decompositions that have been introduced in this work is an open question.

All our fermionic \gls{LCU} decompositions are performed on the two-body part of the Hamiltonian, which we define as the part of the Hamiltonian quartic in Majorana operators. Since we know that the remaining one-body part of the Hamiltonian can be optimally treated using DF \cite{loaiza_lcu}, all 1-norms corresponding to fermionic \gls{LCU} decompositions reported in this paper are, therefore, sum of the spectral range of the one-body part and the 1-norm for the fermionic \gls{LCU} decomposition of the two-body part. We have obtained the \gls{THC} decompositions using a non-linear optimization performed with the Levenberg-Marquardt algorithm \cite{Levenberg,Marquardt}, details of our THC implementation are in Appendix \ref{app:compdet}. The Toffoli gate counts for our THC decompositions are obtained using Openfermion \cite{openfermion} and the reported T-gate counts are 4 times the Toffoli counts \cite{toffoli_t}.

To assess efficiency of various \gls{LCU} decompositions we present 
quantum resource estimates for small molecules (Table~\ref{tab:results}) and 1D hydrogen chains (\fig{fig:linear} and Table~\ref{tab:fits}), computational details are in Appendix \ref{app:compdet}. Most of LCU methods can be partitioned into four categories 
based on the small molecule results: 
1) low T-gate count and medium 1-norms, Pauli and Pauli-OO; 
2) medium 1-norms and medium T-gate count, $L^4$-SVD and THC; 
3) high T-gate count and low 1-norms, AC, OO-AC, and DF; 
4) medium T-gate count and high 1-norms, $L^4$-CP4 and $L^4$-MPS.  

\begin{table*}[t]
\large
\begin{tabular}{|c|c|c|c|c|c|}
\hline
Molecule     & H$_2$       & LiH           & BeH$_2$       & H$_2$O        & NH$_3$        \\ \hline
$\Delta E/2$ & 0.815       & 4.93          & 9.99          & 41.9          & 33.8          \\ \hline
Pauli        & 1.58(632)   & 13.0(1952)    & 22.8(2144)    & 71.9(2512)    & 69.2(5032)    \\ \hline
OO-Pauli     & 1.58(632)   & 12.4(1936)    & 21.9(2136)    & 61.0(2488)    & 54.5(3744)    \\ \hline
AC           & 1.49(1040)  & 10.2(95362)   & 18.0(98522)   & 57.2(166700)  & 48.9(588301)  \\ \hline
OO-AC        & 1.49(1040)  & 10.2(92715)   & 17.9(96860)   & 55.7(162665)  & 46.8(355022)  \\ \hline
DF           & 1.37(3792)  & 9.34(23671)   & 16.4(31529)   & 53.7(31529)   & 44.7(39550)   \\ \hline
THC          & 1.8 (2500)  & 11.18 (5976)  & 19.59 (6832)  & 58.55 (6832)  & 49.26 (8316)  \\ \hline
$L^4$-SVD    & 2.19 (3264) & 11.8 (12086)  & 22.0 (15488)  & 61.1 (15488)  & 53.3 (19548)  \\ \hline
$L^4$-CP4    & 5.65(7556)  & 34.5 (13044)  & 61.1 (14416)  & 125.0 (14416) & 121.0 (15788) \\ \hline
$L^4$-MPS    & 4.63(3750)  & 130.08(12384) & 279.27(14432) & 474.1(14284)  & 612.45(16664) \\ \hline
\end{tabular}
\caption{1-norms for electronic structure Hamiltonians in the STO-3G basis \cite{szabo}, the numbers of T-gates required to implement the associated Hamiltonian oracle circuit are given in parenthesis. $\Delta E/2$ corresponds to a lower bound for the 1-norm. Pauli: Pauli products as unitaries; AC: anti-commuting grouping technique; OO-prefix indicates that orbitals were optimized to minimize the 1-norm; DF: double factorization; THC: tensor hypercontraction; $L^4$-SVD, $L^4$-CP4, and $L^4$-MPS: MTD-$L^4$ decompositions introduced in Sec.~\ref{subsec:mtd_svd}.}
\label{tab:results}
\end{table*}


Overall, the orbital-optimized Pauli \gls{LCU} has the lowest hardness in virtue of the small number of T-gates required for its implementation. The \prep circuit for this decomposition can be optimized to exploit the 8-fold symmetry of the two-electron tensor as shown in Ref.~\citenum{qubitization_df}, greatly reducing the number of coefficients that need to be loaded. This, along with the simple structure of its \sel circuit make it extremely efficient. For the orbital-optimized schemes in small molecules (Table~\ref{tab:results}) we used the full orbital optimization which minimizes the total 1-norm since the Foster-Boys localization scheme did not improve the results. 
In contrast, for 1D hydrogen chains, the orbital rotations that minimize the 1-norm yield results similar to those obtained using the Foster-Boys orbital localization scheme,
which is in line with findings of Ref.~\citenum{majorana_l1}. Therefore, for computational efficiency, we used the Foster-Boys localization scheme for the \glspl{LCU} of hydrogen chains. It is worth noting that fermionic-based decompositions, which inherently include orbital rotations as part of the \gls{LCU}, are unaffected by the orbital basis used to represent the electronic tensor. 

Regarding the scaling of hardness with the system size (Table \ref{tab:fits}), the only decompositions which presented a better scaling than Pauli-OO are $L^4$-CP4 and \gls{THC}. The \gls{THC} decomposition offers the most optimal scaling with increasing system size $N$, hardness scaling of $O(N^{2.19})$ is obtained empirically from results of Table \ref{tab:fits}. Thus, \gls{THC} can be advantageous over Pauli-OO for systems with $N>23$.  We note that extremely large 1-norms were obtained for the $L^4$-CP4 tensor decomposition for the small molecules (Table \ref{tab:results}). We attribute this to an implementation that was used for the CP4 decomposition, and having a more robust implementation will be necessary for larger systems, as discussed in Ref.~\citenum{efficient_cp4}. At the same time, $L^4$-SVD decomposition gave competitive 1-norms for these molecules.  On the other hand, the SVD-based 1-norms for the hydrogen chains were larger than their CP4 counterparts, which makes the SVD decomposition prohibitively expensive for larger chains. Overall, these results establish the $L^4$ \gls{LCU} decomposition as a promising decomposition in terms of hardness; finding the corresponding fragments in a robust and efficient way is work in progress. We believe that this decomposition could be useful for hybrid \gls{LCU} approaches by obtaining a few large MTD-$L^4$ fragments before decomposing the rest of the Hamiltonian with some other method. However, exploring these kind of hybrid \glspl{LCU} is beyond the scope of this work. 

The \gls{AC} decomposition, when using the orbital optimized Hamiltonians, yielded very low 1-norms and requires the fewest qubits. However, the large number of rotations required for its \sel circuit increases the T-gate count and 
makes its hardness one of the worst out of all the studied \glspl{LCU}. As such, we believe that an implementation of the \gls{AC} decomposition which loads the Givens rotation angles in the quantum computer could lower significantly the cost compared to the current implementation that we proposed where angles are ``hard coded'' into the circuits.

The results for the \gls{DF} method correspond to the implementation discussed in the Appendix \ref{app:lcu circuits}. The ``optimized DF'' results in Fig.~\ref{fig:linear} and Table \ref{tab:fits} correspond to the \gls{DF} implementation discussed in Ref.~\citenum{THC} where each Toffoli gate has been considered to require 4 T-gates \cite{toffoli_t}. We note that this decomposition 
also presents one of the best hardness scalings. However, the number of qubits required for the optimized DF is an order of magnitude larger than that required by most other approaches. 
Only the THC decomposition involves similar numbers of ancilla qubits as the Optimized DF approach.

\begin{table}[b!]
    \normalsize
    \centering
    \begin{tabular}{|c|c|c|c|c|} \hline
     LCU method    &  Fitted quantity & $\alpha$ & $\beta$ \\ \hline
    \multirow{2}{*}{Pauli} & $\lambda\times\#$T-gates & 0.01 & 5.59 \\
& \# qubits & 1.49 & 0.44 \\ \hline
\multirow{2}{*}{AC} & $\lambda\times\#$T-gates & 1.85 & 5.53 \\
& \# qubits & 1.35 & 0.45 \\ \hline
\multirow{2}{*}{OO-Pauli} & $\lambda\times\#$T-gates & 2.36 & \textbf{2.67} \\
& \# qubits & 1.54 & 0.41 \\ \hline
\multirow{2}{*}{OO-AC} & $\lambda\times\#$T-gates & 4.04 & 2.82 \\
& \# qubits & 1.39 & 0.41 \\ \hline
\multirow{2}{*}{$L^4$-SVD} & $\lambda\times\#$T-gates & 2.10 & 3.83 \\
& \# qubits & 1.86 & 0.23 \\ \hline
\multirow{2}{*}{$L^4$-CP4} & $\lambda\times\#$T-gates & 3.17 & \textbf{2.36} \\
& \# qubits & 1.85 & 0.23 \\ \hline
\multirow{2}{*}{$L^4$-MPS} & $\lambda\times\#$T-gates & 2.55 & 4.20 \\
& \# qubits & 2.12 & 0.14 \\ \hline
\multirow{2}{*}{DF} & $\lambda\times\#$T-gates & 1.80 & 4.14 \\
& \# qubits & 1.91 & 0.24 \\ \hline
\multirow{2}{*}{Optimized DF} & $\lambda\times\#$T-gates & 2.58 & 2.70 \\
& \# qubits & 0.66 & 1.75 \\ \hline
\multirow{2}{*}{THC} & $\lambda\times\#$T-gates & 3.02 & \textbf{2.19} \\
& \# qubits & 1.62 & 0.91 \\ \hline
    \end{tabular}
    \caption{Linear fit coefficients for Fig.~\ref{fig:linear}, $\log_{10}y = \alpha + \beta \log_{10} x$.}
    \label{tab:fits}
\end{table}

{\it The classical pre-processing cost:} 
The Pauli products decomposition of a Hamiltonian has $\mathcal{O}(N^4)$ operators. However, orbital localization schemes can reduce the scaling to $\mathcal{O}(N^2)$ for large molecules \cite{lmo_qc}; the cost for optimized orbital localization schemes typically scales as $\mathcal{O}(N^3)$ \cite{majorana_l1}. The sorted insertion algorithm, used in \gls{AC}, has a sorting cost of $\mathcal{O}(M^2)$, where $M$ is the number of elements in the list being sorted. This results in a decomposition cost of $\mathcal{O}(N^8)$ for \gls{AC} and $\mathcal{O}(N^4)$ when grouping is done on localized orbitals. For the fermionic methods, the cost for \gls{DF} is dominated by the diagonalization of the reshaped $N^2\times N^2$ tensor $g_{ijkl}$, which scales as $\mathcal{O}(N^6)$. 

\onecolumngrid
 \begin{figure*}[t!]
\centering
\includegraphics[width=16cm]{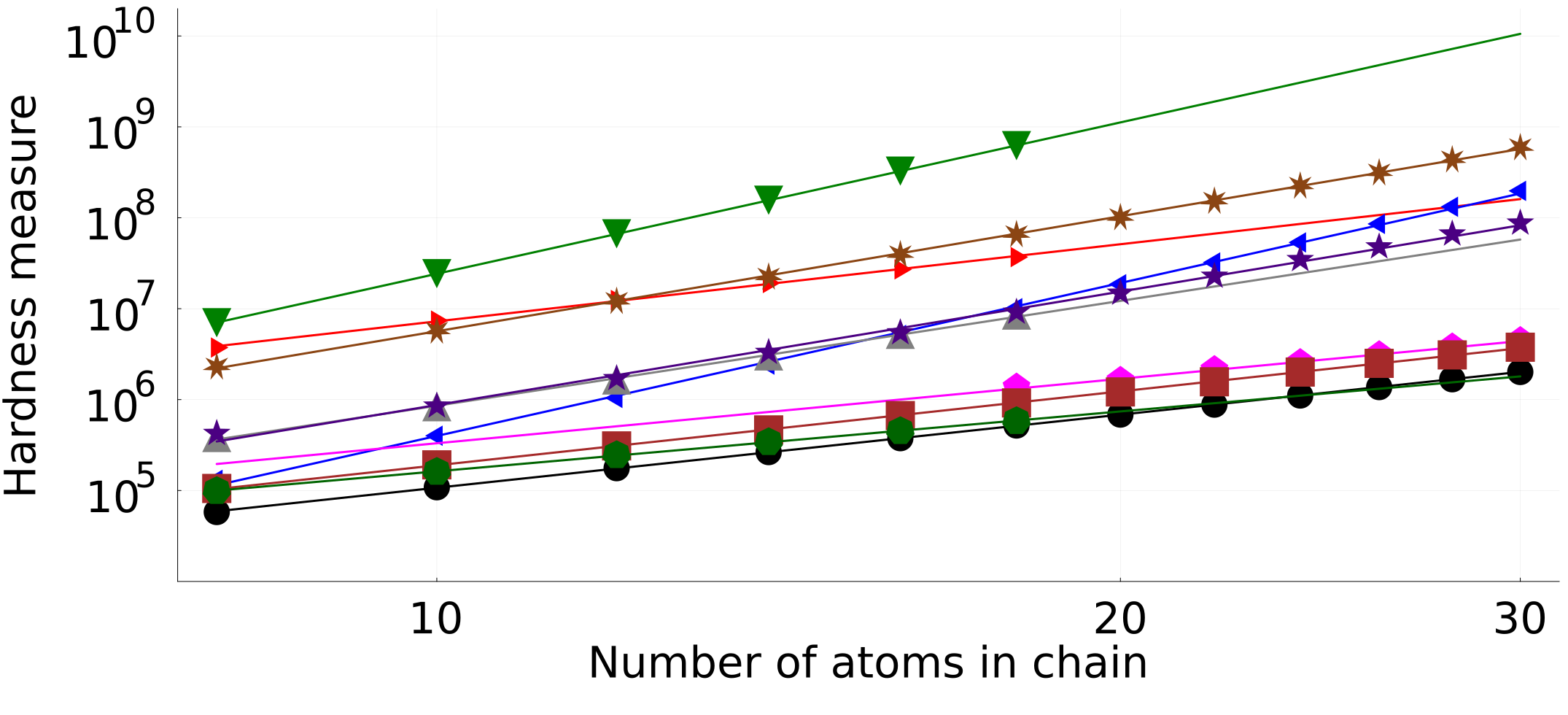}
\includegraphics[width=16cm]{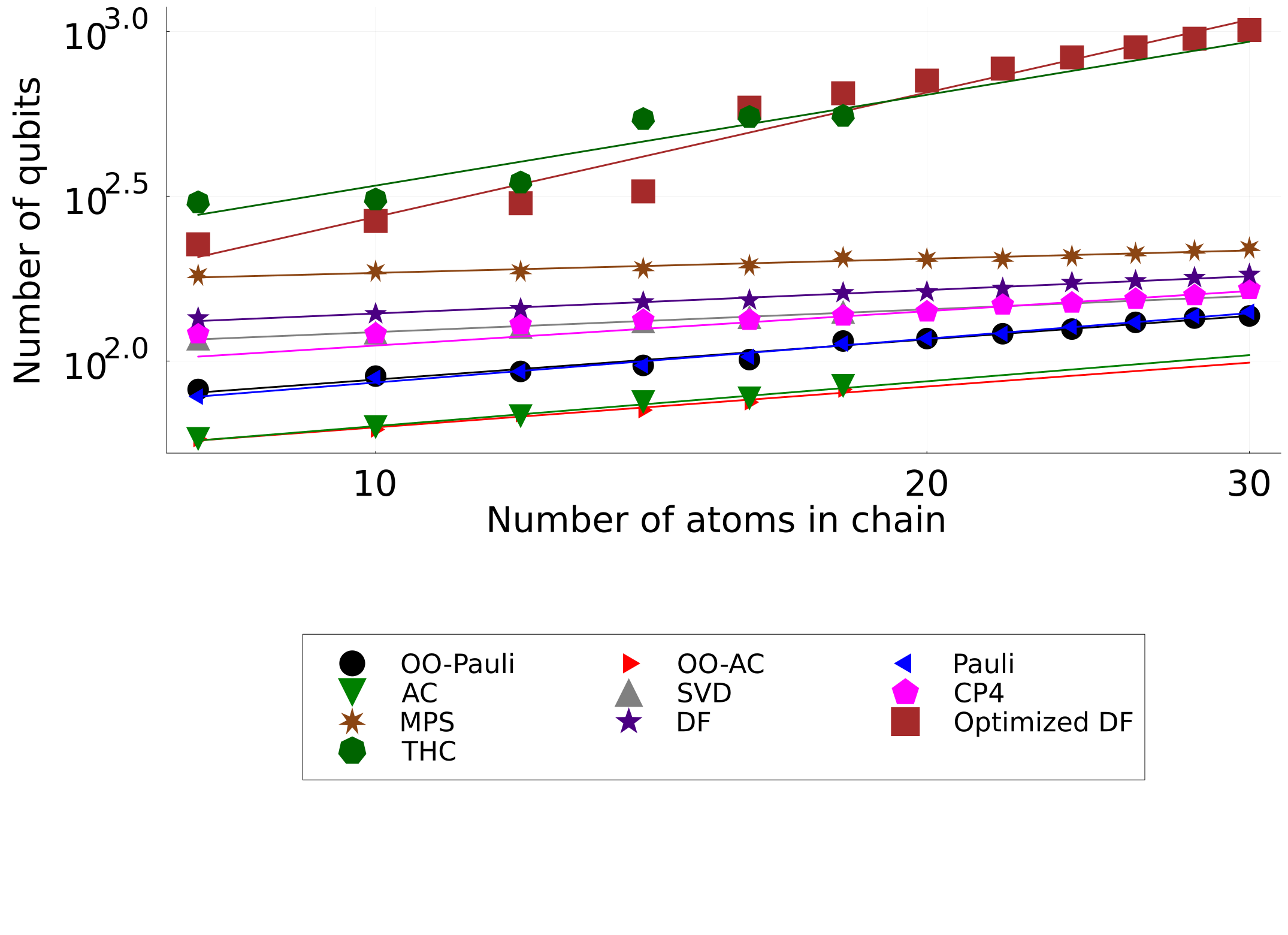}
\vspace{-1cm}
\caption{Hardness (top) and the number of qubits (bottom) scaling for different \gls{LCU}s for hydrogen chains with $N$ atoms with a spacing $r=1.4$ \r{A}. Hardness is defined as the product of the 1-norm $\lambda$ with the number of T-gates for one \sel and two \prep oracles. Lines correspond to a least-square fit, $\log_{10}y = \alpha + \beta \log_{10} x$. Coefficients $\alpha$ and $\beta$ are reported in Table~\ref{tab:fits}.  Orbital optimizations were done using the Foster-Boys localization \cite{FB}.} \label{fig:linear}
\end{figure*}

\newpage
\mbox{}
\newpage
\twocolumngrid

For our \gls{THC} implementation, the computation of an approximate Hessian matrix for the update step in the Levenberg-Marquardt algorithm dominates the computational cost. Assuming that the rank of the THC decomposition, $K$ [\eq{thc}] scales linearly with system size, the computational cost of our THC implementation scales as $O(N^8)$. This is in contrast to the $O(N^6)$ scaling of Openfermion's L-BFGS-B THC implementation. Our implementation of THC offers greater robustness with the choice of initial parameters in exchange for the higher computational scaling. 
For $L^4$-MPS, the cost for each step in the \gls{MPS} factorization (see Fig.~\ref{fig:SVD_MPS}) goes as $\mathcal{O}(N^5)$, thus being the overall scaling cost of this decomposition. The $L^4$-SVD decomposition quickly becomes prohibitively expensive to do as the system size grows, while the $L^4$-CP4 decomposition requires $\mathcal{O}(N^3 W)$, where $W$ is the number of terms in the decomposition \cite{efficient_cp4}. Among the algorithms presented in this work, orbital-localized \gls{AC} has the best theoretical scaling, followed by $L^4$-MPS. However, in practice, the fermionic-based decompositions (\gls{DF} and $L^4$-MPS) tend to have lower prefactors due to their efficient utilization of linear algebra routines. For example, in the case of the linear H$_{30}$ chain, orbital-localized \gls{AC} took approximately 5 minutes, while \gls{DF} required around 0.1 seconds and $L^4$--MPS required around 0.3 seconds. These calculations were done using a single core of a 2.2 GHz 6-Core Intel i7 processor. 

\newpage
\section{Conclusion}

In summary, this work has introduced the \gls{MTD} framework, which provides a unified approach to \gls{LCU} decompositions of the electronic structure Hamiltonian. This framework not only presents a unifying perspective but also leads to new decompositions like \gls{MTD} $L^4$, the latter expresses the electronic structure tensor as a minimal-rank representation. 
\gls{MTD} establishes a connection between popular tensor decompositions and \glspl{LCU}. A study over small systems revealed the \gls{LCU} based on Pauli products with an orbital optimization to be the most competitive. However, the MTD $L^4$-CP4 and \gls{THC} decompositions had better scalings with the system size, suggesting that for larger systems these decompositions could be more efficient to implement.

\section*{Acknowledgements}
We are grateful to Nathan Wiebe and Nicholas Rubin for insightful discussions, and Alexey Uvarov, Sangeeth D. Kallullathil, Shreyas Malpathak, and Joshua T. Cantin for valuable comments on the manuscript. I.L. and A.F.I. gratefully appreciate financial support from the Mitacs Elevate Postdoctoral Fellowship and Zapata Computing Inc. 

\appendix
\onecolumngrid

\section{LCU circuits}\label{app:lcu circuits}

In this section we give an overview of all the necessary components for the \prep and \sel circuits for the different \glspl{LCU}. Most of these circuits have been developed in Refs.\citenum{qrom,THC,femoco_df}, here, we include circuits that have not been explicitly drawn before, along with those corresponding to the newly proposed \glspl{LCU} in this work. We make extensive use of the QROM and unary iteration circuits shown in Ref.~\citenum{qrom}, these procedures are denoted on our circuit diagrams as ``$In$'' and ``$\textrm{data}$'' directives, respectively. Note that usage of a more advanced version of QROM, i.e. QROAM \cite{qroam}, could improve the circuits by modifying the T-gate and ancilla counts. However studying the additional benefits of this technique are beyond the scope of this work. In all the shown circuits, we use the convention where having a single-qubit gate (e.g., Hadamard) applied to a multi-qubit register corresponds to applying that gate on each of the qubits of the register. An overview of the resource cost for all circuits is given in Table~\ref{tab:resources}, with pertinent parameters defined either in the Table~\ref{tab:resources} caption or in the associated circuit in this section.

\begin{sidewaystable}[]
    \centering
        \caption{T-gate, number of logical qubits, and number of controlled $R_Z$ rotations cost of quantum circuits of interest for this work. Reusable qubits corresponds to those that are left in the $\ket{0}$ state after the directive is executed, and is also included in the logical qubit count. We have used $b_n \equiv \myceil{\log_2 n}$, $k_n \equiv \myfloor{\log_2 n}$, and $l_n \equiv \myceil{\log_2 (n/(2^{k_n}))}$ such that $b_n = k_n + l_n$. $\mu_n \equiv \myceil{\log_2(n\epsilon)}$ corresponds to the number of bits required for the ``alt'' register used when preparing a coherent superposition of $n$ coefficients with an accuracy of $\epsilon$. In general, $\epsilon$ corresponds to the target accuracy for the given directive and $\lambda$ to the 1-norm of the LCU. $N$ represents the number of spacial orbitals. Numbers in braces for \prep circuit correspond to additional resources when circuits are controlled by an additional qubit, shown by the $\ket{ctl}$ register in the corresponding figures. We always consider all \sel operations to be controlled by a $\ket{ctl}$ qubit. All constants are defined in the caption of the corresponding figure. As shown in Ref.~\citenum{optimal_t}, $R_Z$ rotations can be optimally implemented for a target accuracy $\epsilon<0.016$ with an average T-gate cost of $3.067*\log_2\left( 1/\epsilon \right) + 9.678$ and one additional reusable qubit. The controlled Givens$(N)$ directive corresponds to an orbital rotation consisting of $N-1$ Givens rotations that are controlled by a register where the corresponding angles have been previously loaded, as shown in Refs.~\citenum{femoco_df,THC}. The rotations for this directive can be implemented using the phase gradient technique \cite{phase_grad} for a cost no larger than $7$ T-gates per rotation. Note that our notation for $N$ differs by a factor of $2$ to that in Ref.~\citenum{THC}.}
    \label{tab:resources}
    \begin{adjustbox}{scale=0.95,center}
    \begin{tabular}{|c|p{5.7cm}|p{5.7cm}|c|c|}
    \hline
        Circuit directive & T-gates complexity $(\#T)$ & Number of non reusable qubits $(\#Q)$ & Number of reusable qubits $(\#R_Q)$ &  $R_Z$ complexity $(\#R_Z)$ \\ \hline \hline
        UNIFORM$(K)$ (Ref.~\cite{qrom}) & $8l_K + \{2l_K + 2k_K\}$ & $k_K + l_K + \{1\}$ & $l_K$ & $2$ \\ \hline
        Controlled SWAP$(b\leftrightarrow b)$ & $7b$ & $2b+1$ & $0$ & $0$ \\ \hline
        Controlled Givens$(N)$ & \parbox{6.5cm}{$14N(\beta-2)$ \\ $\beta \equiv \myceil{5.652 + \log_2 N\lambda/\epsilon}$} & $N + \beta + 1$& $0$ & $0$ \\ \hline
        $\mathtt{PREP}(K)$ (Fig.~\ref{fig:gen_prep})& $8l_K + 4K + 8\mu_K + 7b_K - 8 + \{4 + 2k_K + 2l_K\}$ & $b_K + 2\mu_K + 3 +\{1\}$ & $+ \max\left[2\mu_K-1; b_K-1; l_K\right] + \{1\}$ & $2$ \\ \hline
        $\mathtt{PREP}_V(K)$ (Fig.~\ref{fig:V_prep}) & $\#T(\mathtt{PREP}(K)) + 7$ & $\#Q(\mathtt{PREP}(K)) + 2$ & $\#R_Q(\mathtt{PREP}(K))$ & $2$ \\ \hline
        $\mathtt{PREP}({\rm Sparse})$ (Fig.~\ref{fig:sparse_prep}) & $8l_S + 4S + 8\mu_S + 56b_N - 1$ & $b_S + 8b_N + 2\mu_S + 8$ &  $\max[l_S; b_S-1; 2\mu_S - 1]$ & $2$ \\ \hline
        $\mathtt{SEL}({\rm Sparse})$ (Fig.~\ref{fig:sel_sparse}) & $32N - 16$ & $4b_N + 4 + 2N$ & $b_{2N} + 1$ & $0$ \\ \hline
        $\mathtt{PREP}({\rm AC})$ & $\#T(\mathtt{PREP}(G))$ & $\#Q(\mathtt{PREP}(G))$ & $\#R_Q(\mathtt{PREP}(G))$ & $2$ \\ \hline
        $\mathtt{SEL}({\rm AC})$ (Fig.~\ref{fig:sel_AC}) & $4G-4$ & $2N + b_G + 1$ & $b_G$ & $2\left(\sum_n G_n\right) - 2G$ \\ \hline
        $\mathtt{PREP}(L^4)$& $\#T(\mathtt{PREP}(W))$ & $\#Q(\mathtt{PREP}(W))$ & $\#R_Q(\mathtt{PREP}(W))$ & $2$ \\ \hline
        $\mathtt{SEL}(L^4)$ (Fig.~\ref{fig:sel_CP4T})& $N(112\beta-196) + 8W - 4$ & $4 + b_W + 2N + 4\beta$ & $b_W$ & 0 \\ \hline
        $\mathtt{PREP}(L^4 {\rm -MPS})$ (Fig.~\ref{fig:prep_MPS}) & $\#T(c\mathtt{PREP}(N)) + \#T(c\mathtt{PREP}(\alpha_1)) + \#T(c\mathtt{PREP}(\alpha_2)) + \#T(c\mathtt{PREP}(\alpha_3))$ & $13 + b_N + b_{\alpha_2} + b_{\alpha_3} + 2\mu_{N} + 2\mu_{\alpha_1} + 2\mu_{\alpha_2} + 2\mu_{\alpha_3}$ & $4 + b_{\alpha_2} + 2\mu_{\alpha_2}$ & $9$ \\ \hline
        $\mathtt{SEL}(L^4{\rm -MPS})$ (Fig.~\ref{fig:sel_MPS})& $4\alpha_2(N + 2\alpha_1 + 2\alpha_3) + N(112\beta-192) + 8\alpha_1 + 4\alpha_3 - 24$ & $4 + 2N + \beta + b_N + b_{\alpha_2} + b_{\alpha_3}$ & $b_{\alpha_2\alpha_3}$ & 0 \\ \hline 
        $\mathtt{PREP}(\rm{DF}) (Fig.~\ref{fig:DF})$ &  $L(8+40l_N+16N+32\mu_N+28b_N+8k_N) + N(28\beta-48) + 4b_N - 20$ & $6+b_L+2N+b_N+2\mu_N$ & $7+\beta+3b_N+b_L+\max[2\mu_N-1;b_N-1;l_N]$ & $8L$ \\ \hline
    \end{tabular}
\end{adjustbox}
\end{sidewaystable}

\subsection{\prep circuits}

\begin{figure*}
    \centering
    \includegraphics[scale=0.25]{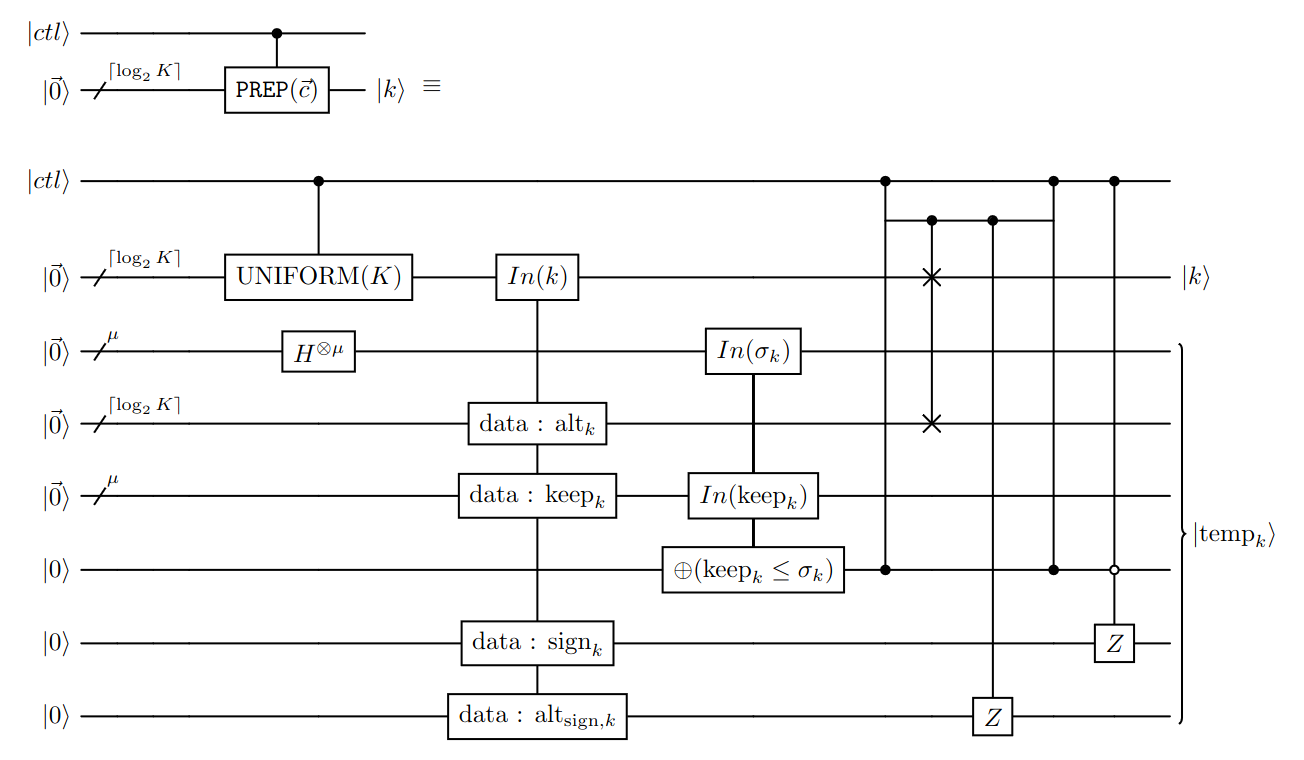}
    \caption{Generic \prep: controlled coherent state preparation circuit for a coefficient vector $\vec c \equiv (c_0, c_1, ..., c_{K-1})$, as shown in a non-controlled version in Fig. 11 of Ref.~\citenum{qrom}.  The coefficients $c_k$ are implemented with a $\mu$-bit approximation $\tilde c_k$ such that $|c_k - \tilde c_k| \leq 1/2^\mu K$. For calculating the complexity of this circuit, we have defined $l$ and $L$ such that $K=2^l L$ where $L$ is an odd number different than $1$, noting that for $L=1$ we only need to apply Hadamard gates to implement the uniform superposition. The alt and keep values can be obtained from specifying $\vec c$ and $\mu$, as shown in Ref.~\citenum{qrom}.}
    \label{fig:gen_prep}
\end{figure*}

\begin{figure*}
    \includegraphics[scale=0.25]{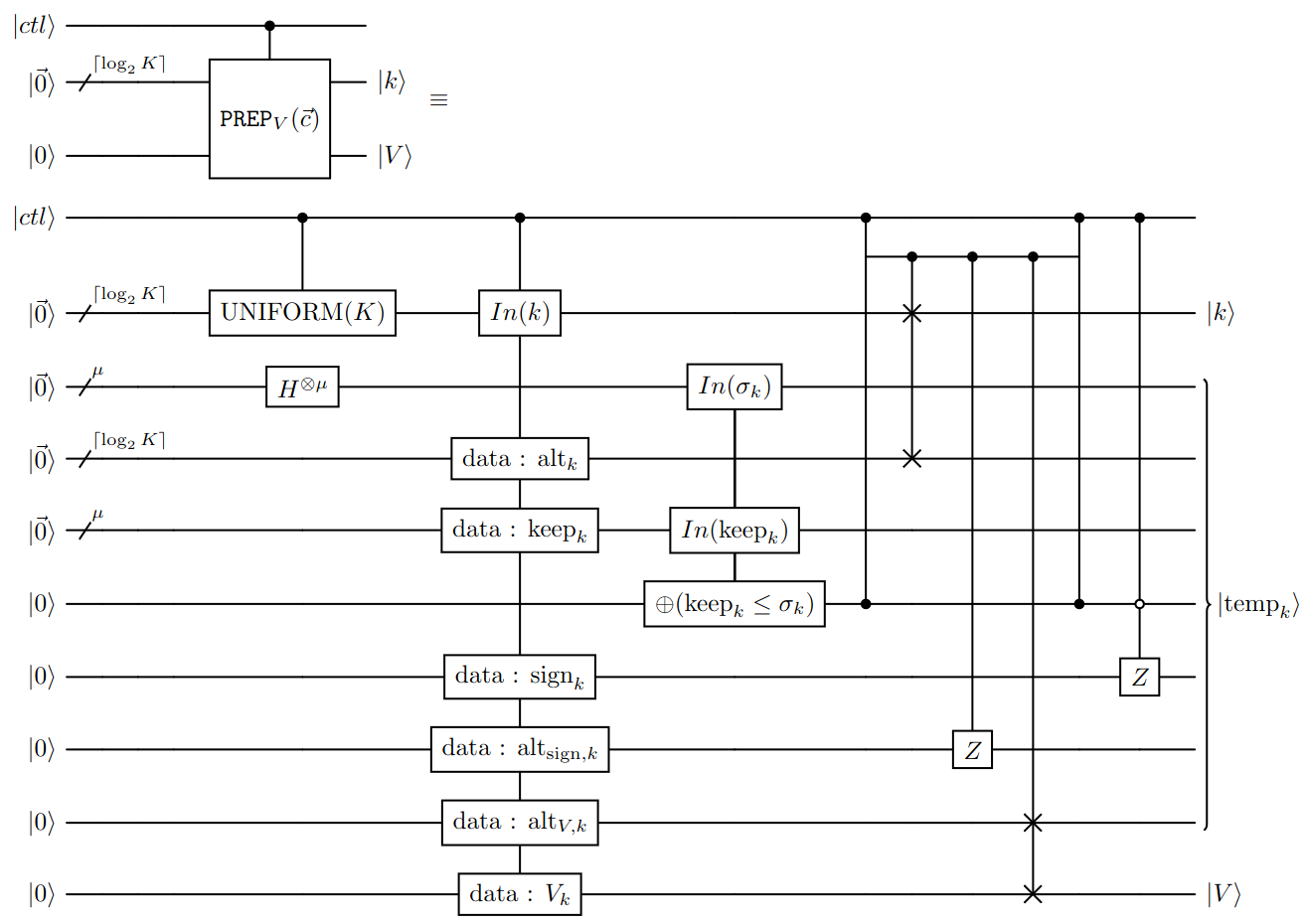}
\caption{Coherent state preparation that also prepares a qubit register $V$ for flagging a particular set of coefficients, e.g., one-electron coefficients.}
\label{fig:V_prep}
\end{figure*}

Here we discuss the \prep circuits for all \glspl{LCU} shown in this work that have not been explicitly drawn in the past. The circuit labeled as UNIFORM$(K)$ will be used throughout, it  prepares an equal superposition over $K$ states with accuracy $\epsilon$, and is shown in Fig. 12 of Ref.\citenum{qrom}. We have also included the sign preparation routine in these circuits, making the output prepare the coefficients with their corresponding signs. A general \prep circuit for an arbitrary coefficient is shown in Fig.~\ref{fig:gen_prep}, which can be used to implement the AC \prep oracle. A \prep circuit that additionally prepares a $V$ register that is used for flagging the one-electron terms, is shown in Fig.~\ref{fig:V_prep}; this circuit can be used for the CP4 decomposition. Circuits for THC are discussed in Ref.~\citenum{THC}. 


The quantum resource components for the generic \prep (Fig.~\ref{fig:gen_prep}) are:
\begin{itemize}
    \item uniform superposition over $K$ states, using two R$_Z$'s, and $8l_K$ T-gates, where $K = l_K 2^{k_K}$ for the largest possible integer $k_K$ such that $l_K \geq 1$ (This operation requires $l_K$ reusable qubits.)
    \item QROM over $K$ indices, requiring $4(K-1)$ T-gates and $b_K - 1$ reusable qubits
    \item arithmetic gate over the keep and $\sigma$ registers, requiring $4(2\mu_K-1)$ T-gates and $2\mu_K - 1$ reusable qubits
    \item one controlled swap gate on the prepare and alt registers, requiring $7b_K$ T-gates
    \item Hadamard and controlled Z operations with no fault-tolerant cost contributions
    \item if the circuit is applied in a controlled manner, the uniform superposition uses an additional $2l_K + 2k_K$ number of T-gates;  we also require one additional reusable qubit, and $4$ extra T-gates for the register with the $And$ operation between the additional control register and the arithmetic operation register.
\end{itemize}
Considering all of these operations, the total cost parts for the generic \prep circuit are:
\begin{itemize}
    \item $8l_K + 4(K-1) + 4(2\mu_K - 1) + 7b_K = 8l_K + 4K + 8\mu_K + 7b_K - 8$ T-gates
    \item $b_K + 2\mu + 3$ non-reusable qubits
    \item $\max[l_K,b_K-1,2\mu_K-1]$ reusable qubits
    \item $2$ $R_Z$ rotations
    \item additional $2l_K + 2k_K + 4$ T-gates, one non-reusable qubit, and one reusable qubit, if the operation is controlled.
\end{itemize}
For the generic \prep where we also have the $V$ register, the cost is the same as for the generic \prep, with two additions:  
\begin{itemize}
    \item one controlled swap for a cost of $7$ T-gates
    \item two additional non-reusable qubits.
\end{itemize}

\subsubsection*{Sparse Pauli}

\begin{figure*}
    \centering
    \includegraphics[scale=0.3]{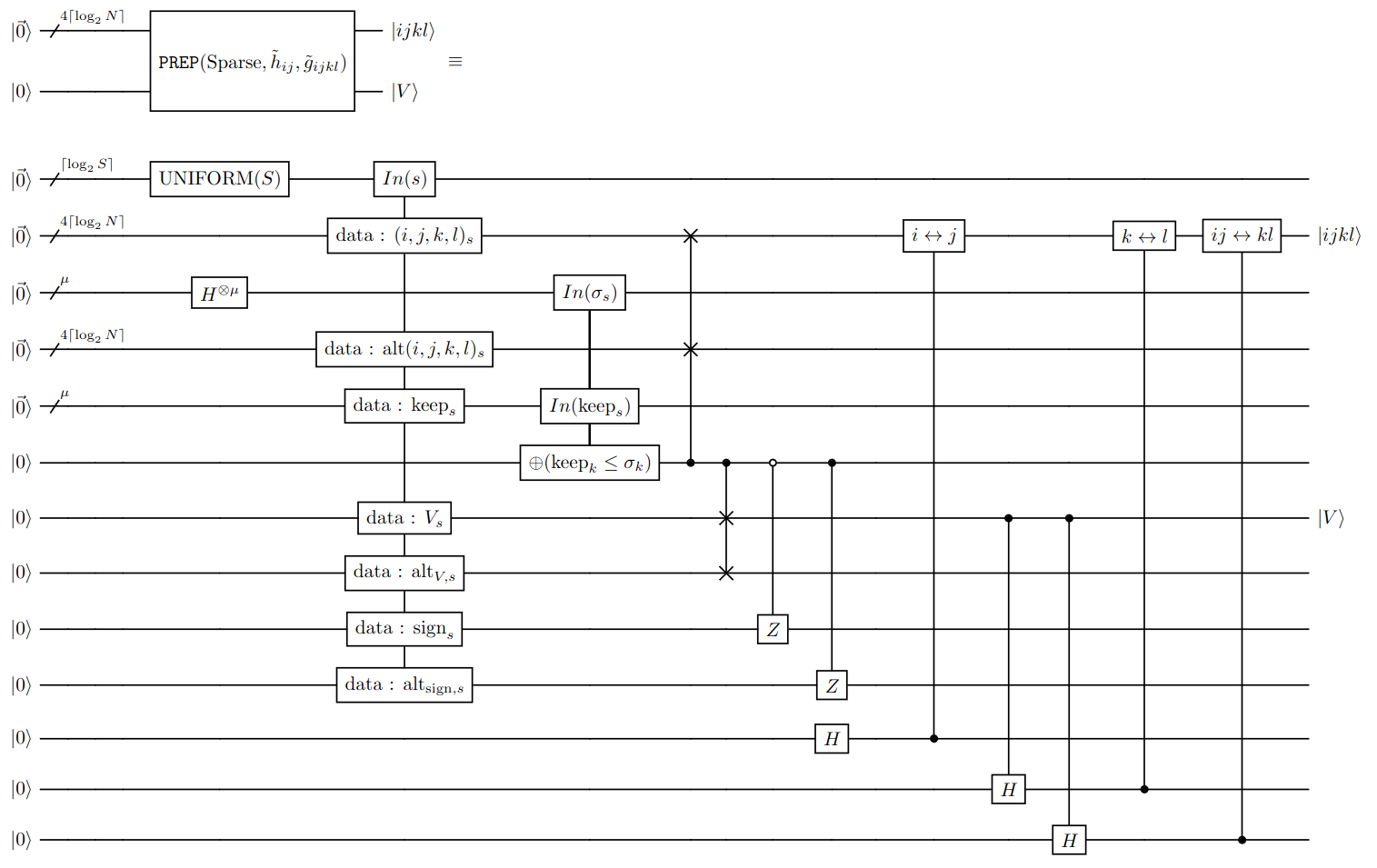}
    \caption{Sparse state preparation circuit for coefficient tensors $h_{ij}$ and $\tilde g_{ijkl}$, as discussed in Ref.~\citenum{qubitization_df}. We create a coefficient vector with length $S$ and values corresponding to the non-zero entries of $\zeta_{ij}h_{ij}$ and $\zeta_{ij}\zeta_{kl}\zeta_{ijkl}\tilde g_{ijkl}$ for defining $\textrm{keep}_s$, which also defines the set of indices $(V,i,j,k,l)_s$ that are used for the index-loading and alt routines. Thus, only non-zero entries are loaded by QROM, along with the corresponding indices for $i,j,k,l$, and $V$ which indicates if the coefficient comes from $h_{ij}$ or $\tilde g_{ijkl}$ for $V=0,1$ respectively. The controlled swap operations in the $i,j,k,l$ register swap between these registers as respectively indicated by the parenthesis on top of the crosses.}
    \label{fig:sparse_prep}
\end{figure*}

The \prep circuit for the Pauli LCU can be implemented efficiently by exploiting the sparsity of the $g_{ijkl}$ tensor \cite{qubitization_df}. The corresponding circuit is shown in Fig.~\ref{fig:sparse_prep}. The symmetry of the electronic tensors has been used for reducing the amount of data to load, having defined the auxiliary variables
\begin{align}
    \zeta_{ij} &\equiv \begin{cases}
    \sqrt{2},\hspace{0.7cm} i<j, \\
    1,\hspace{1cm} i=j, \\
    0,\hspace{1cm} i>j,
    \end{cases}
    \\
    \zeta_{ijkl} &\equiv \begin{cases}
\sqrt{2},\hspace{0.7cm} i<k\ \mathrm{ or }\ i=k\ \mathrm{ and }\ j<l\\
1,\hspace{1cm} i=k\ \mathrm{ and }\ j=l\\
0,\hspace{1cm} i>k\ \mathrm{ or }\ i=k\ \mathrm{ and }\ j>l
\end{cases} \\
\tilde g_{ijkl} &\equiv \begin{cases}
    1,\hspace{1cm} |g_{ijkl}| \geq \epsilon_{\textrm{sparse}}, \\
    0,\hspace{1cm} |g_{ijkl}| < \epsilon_{\textrm{sparse}},
\end{cases}
\end{align}
where we have judiciously assigned the tolerance $\epsilon_{\rm{sparse}} = 1\times 10^{-5}$. By using these definitions, only $1/2$ and $1/8$ of the coefficients need to be loaded on the quantum computer, for the one- and two-electron tensors respectively.

The circuit quantum resources for the sparse \prep circuit are:
\begin{itemize}
    \item uniform superposition over $S$ coefficients, requiring 2 R$_Z$, $8l_S$ T-gates, and $l_S$ reusable qubits
    \item QROM over $S$ indices, requiring $4(S-1)$ T-gates and $b_S - 1$ reusable qubits
    \item arithmetic gate, requiring $4(2\mu_S - 1)$ T-gates and $2\mu_S - 1$ reusable qubits
    \item controlled swaps, requiring in total $7(4b_N + 1 + b_N + b_N + 2b_N) = 56b_N + 7$ T-gates
    \item controlled Z's and Hadamard gates, for no additional fault-tolerant cost.
\end{itemize}

The total cost components of the sparse \prep circuit then becomes
\begin{itemize}
    \item $8l_S + 4(S-1) + 4(2\mu_S - 1) + 56b_N + 7 = 8l_S + 4S + 8\mu_S + 56b_N - 1$ T-gates.
    \item $b_S + 8b_N + 2\mu_S + 8$ non-reusable qubits.
    \item $\max[l_S, b_S-1, 2\mu_S-1]$ reusable qubits.
    \item $2$ $R_Z$ rotations.
\end{itemize}

\subsubsection{Matrix Product States}

\begin{figure}
    \centering
    \includegraphics[scale=0.25]{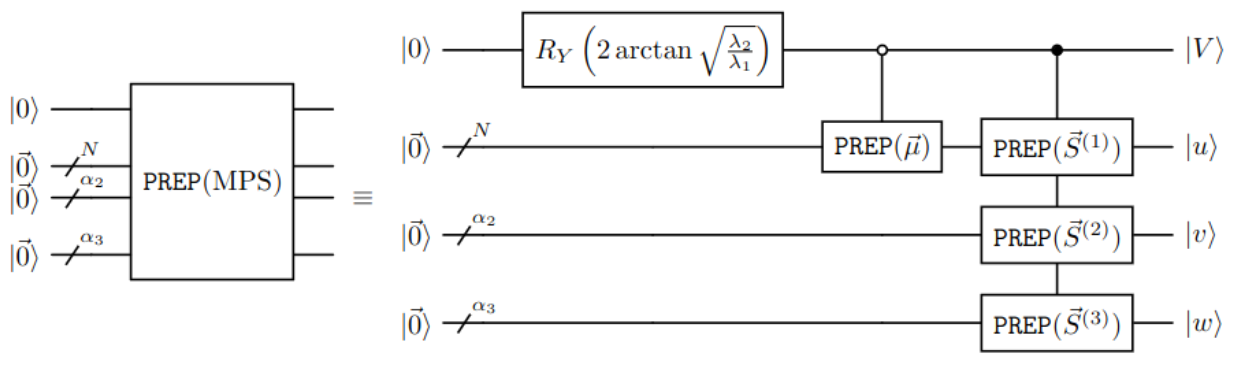}
    \caption{\prep circuit for MPS decomposition. $\lambda_1$ and $\lambda_2$ correspond respectively to the 1-norm of the one- and two-electron components. $\vec\mu$ corresponds to the vector with the components of the one-electron operator after diagonalization. The $\vec S^{(i)}$ coefficients are obtained with the nested SVD for obtaining the MPS, as shown in Eq.~\eqref{eq:mps}}.
    \label{fig:prep_MPS}
\end{figure}

The structure of the MPS decomposition makes it amenable to a more optimal \prep circuit than that of the $L^4$ decomposition, and is shown in Fig.~\ref{fig:prep_MPS}. 
The associated circuit quantum resources are:
\begin{itemize}
    \item one $R_Z$ rotation gate
    \item one controlled generic \prep circuit over $N$ indices
    \item three controlled generic \prep circuits over respectively $\alpha_1$, $\alpha_2$, and $\alpha_3$ indices.
\end{itemize}
The total cost for this circuit parts are:
\begin{itemize}
    \item $\sum_A \left(10 l_A + 2k_A + 4A + 8\mu_A + 7b_A - 4\right)$ T-gates, where $A = N,\alpha_1,\alpha_2,\alpha_3$
    \item $1 + b_N + b_{\alpha_2} + b_{\alpha_3} + \sum_A (2\mu_A + 3)$ non-reusable qubits
    \item $4 + b_{\alpha_2} + 2\mu_{\alpha_2}$ reusable qubits
    \item $9$ $R_Z$ rotations.
\end{itemize}

\subsection{\sel circuits}

We now illustrate how the \sel circuits of the different LCUs can be implemented. The circuits for the THC decomposition and the most optimized version of DF are shown in Ref.~\citenum{THC}. We note that all \sel circuits shown here are controlled by a control register, as required for the implementation of the qubitized walk operator. All required inputs for these circuits are given as an output by the \prep oracles. The $\ket{\psi}$ register encodes the electronic wavefunction over all $2N$ spin orbitals, while $\ket{\psi_\uparrow}$/$\ket{\psi_\downarrow}$ corresponds to the register with only up/down spins and thus having $N$ qubits.

\subsubsection{Sparse Pauli}
\begin{figure*}
    \centering
    \includegraphics[scale=0.3]{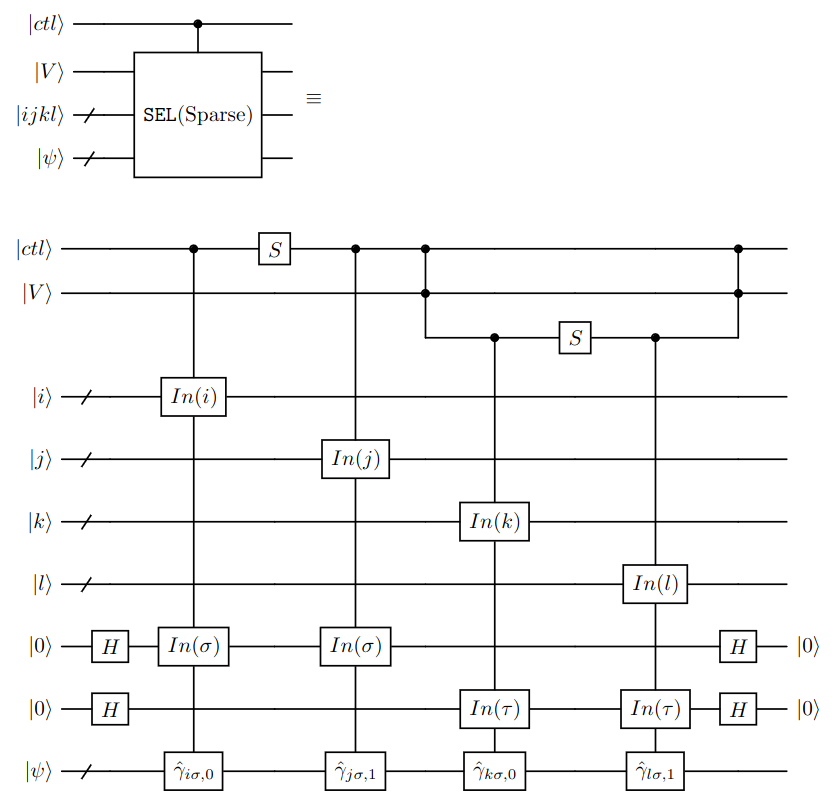}
    \caption{Controlled \sel circuit for Pauli (or sparse) decomposition. An optimal implementation of the multiplexed Majorana operators is shown in Ref.~\citenum{qrom}.}
    \label{fig:sel_sparse}
\end{figure*}

The \sel circuit for the sparse LCU is shown in Fig.~\ref{fig:sel_sparse}. The circuit complexity for this \sel circuit is:
\begin{itemize}
    \item $4$ controlled Majorana operations over $2N$ indices, each one requiring $4(2N-1)$ T-gates and $b_{2N}$ reusable qubits.
    \item One $And$ operation, requiring $4$ T-gates and one reusable qubit.
    \item All the Hadamards and S-gates can be implemented without a fault-tolerant cost.
\end{itemize}
The total cost of the circuit then becomes

\begin{itemize}
    \item $32N - 16$ T-gates.
    \item $4b_N + 4 + 2N$ non-reusable qubits.
    \item $b_{2N} + 1$ reusable qubits.
\end{itemize}

\subsubsection{Anti-commuting groups of Pauli products}
\begin{figure*}
    \centering
    \includegraphics[scale=0.25]{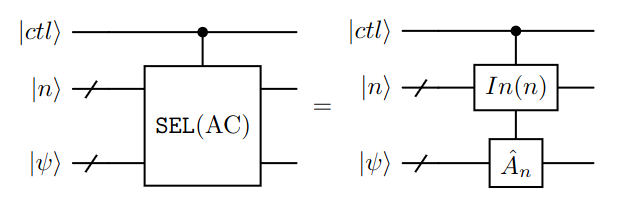}
    \caption{\sel oracle for the AC LCU, which used unary iteration over $n$ index. The circuit for each controlled $\hat A_n$ unitary is shown in Fig.~\ref{fig:An}.}
    \label{fig:sel_AC}
\end{figure*}

The Givens-based diagonalization for implementing the unitaries coming from the AC grouping is discussed in Sec.~\ref{subsec:AC}. The \sel circuit for AC is then shown in Fig.~\ref{fig:sel_AC}. The associated cost for this circuit is
\begin{itemize}
    \item one controlled unary iteration over $G$ indices, requiring $4(G-1)$ T-gates and $b_G$ reusable qubits
    \item $G$ controlled $\hat A_n$ operators, each one requiring two Givens diagonalization circuits over $G_n$ terms and a controlled Pauli with no fault-tolerant cost. (Each Givens diagonalization circuit requires $G_n-1$ controlled R$_Z$ when using the staircase algorithm for implementing exponentials of Pauli products \cite{staircase}.) 
\end{itemize}
The total cost for this \sel circuit is:
\begin{itemize}
    \item $4G-4$ T-gates
    \item $b_G + 1 + 2N$ non reusable qubits
    \item $b_G$ reusable qubits
    \item $2\left(\sum_n G_n\right) - 2G$ controlled $R_Z$ rotations.
\end{itemize}

\begin{figure*}
    \centering
    \includegraphics[scale=0.25]{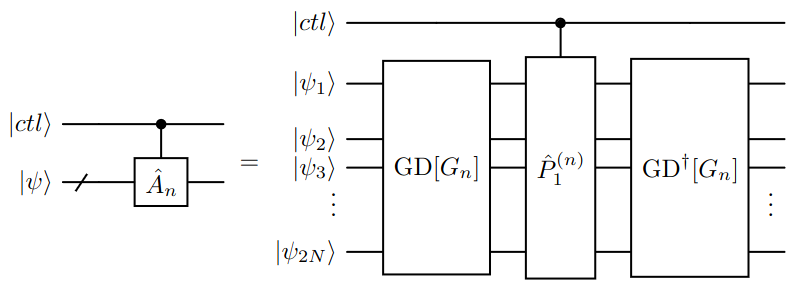} 
   \\ \includegraphics[scale=0.25]{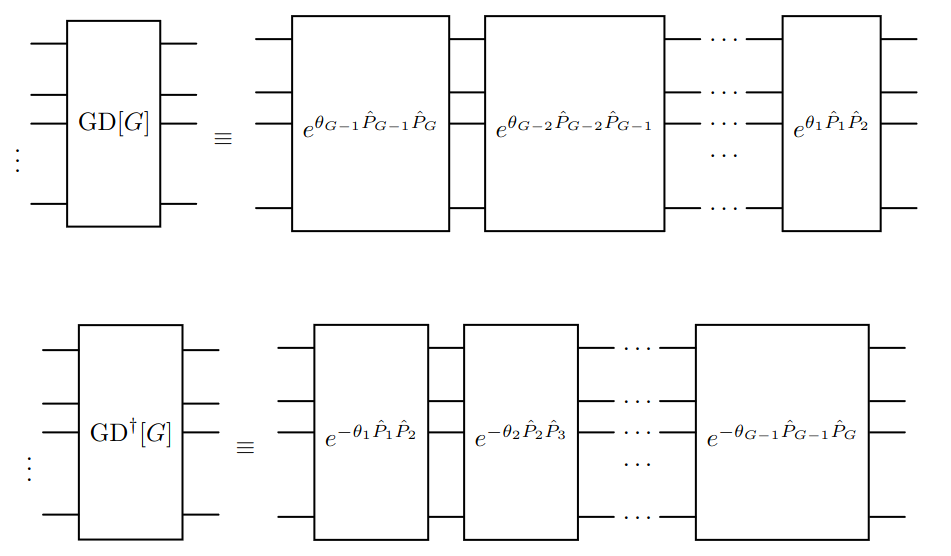}    \caption{Circuit for implementing a controlled $\hat A_n$ unitary for the AC decomposition (top). The associated Givens diagonalization unitaries are shown in the middle and in the bottom. Here, $G_n$ corresponds to the $n$-th group of mutually AC Pauli products, with $P_k^{(n)}$ its $k$-th element.}
    \label{fig:An}
\end{figure*}

\subsubsection{Double factorization}

The double factorization circuit can be implemented in different ways. 
First proposed in Ref.~\citenum{femoco_df}, a more efficient implementation was then proposed in Ref.~\citenum{THC}. However, the resulting circuits for the latter are significantly harder to analyze. As such, in this work we propose a hybrid implementation, which is based on that of Ref.~\citenum{femoco_df} with the only difference that both one- and two-electron terms are implemented in the same circuit by having a qubit flagging the one-electron part, as done in Ref.~\citenum{THC}.
\begin{figure*}
    \centering
        \includegraphics[scale=0.3]{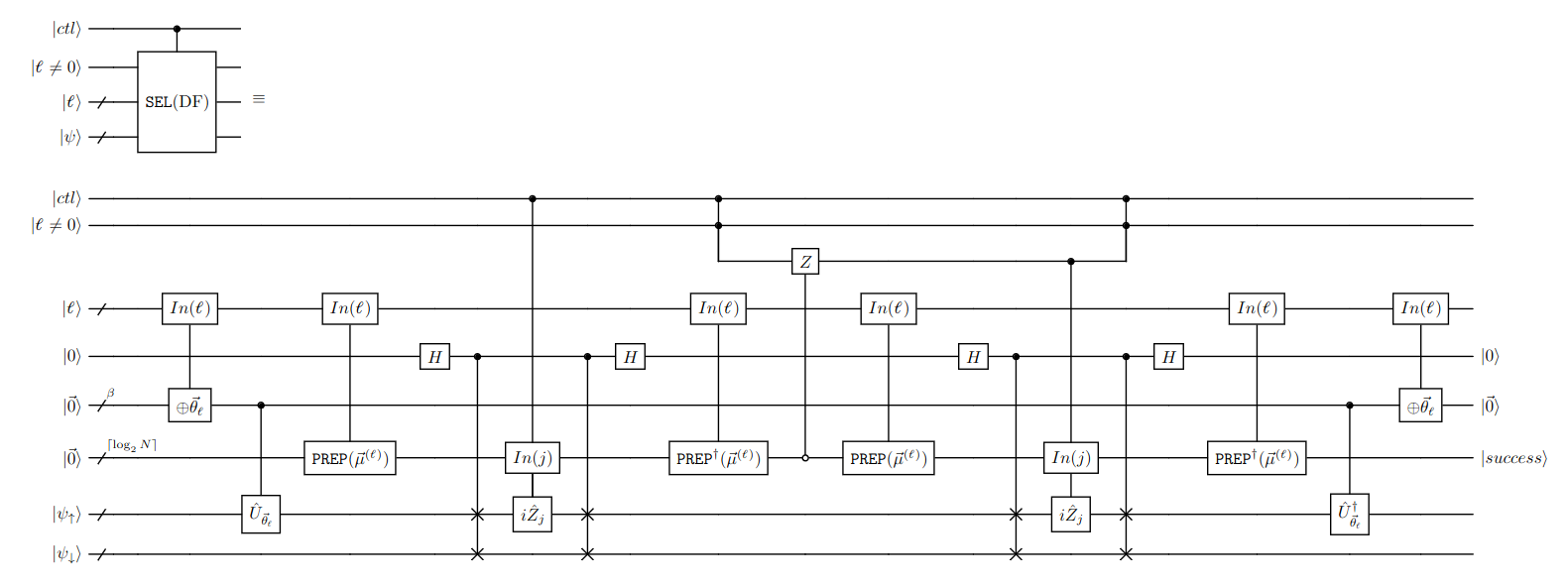}
        \caption{Circuit for implementing a \sel oracle of double factorized LCU. Note how each of the unitaries depending on $\ket{\mathcal{l}}$ is implemented by block-encoding. The register with the $\ket{success}$ encodes whether the application of these block-encodings was successful.}
    \label{fig:DF}
\end{figure*}
The associated cost components to implement this circuit are:
\begin{itemize}
    \item $6$ QROM circuits over $L$ indices, requiring $24(L-1)$ T-gates and $b_{L}$ reusable qubits, here $L=M_{DF}+1$
    \item $2$ multiplexed Givens rotation circuits, each one requiring $14N(\beta-2)$ T-gates
    \item  $4L$ controlled \prep circuits over $N$ coefficients, each one requiring $10l_N + 4N + 8\mu_N + 7b_N - 4 + 2k_N$ T-gates, $b_N + 2\mu_N + 4$ non-reusable qubits, $\max[2\mu_N-1;b_N-1;l_N]+1$ reusable qubits, and $2$ $R_Z$ rotations (Note that the non-reusable qubits can be shared for all these \prep circuits)
    \item $4$ Hadamard gates, for no fault-tolerant cost
    \item $2$ controlled QROMs over $N$ indices, requiring $8N$ T-gates and $b_{N}+1$ reusable qubits
    \item one $Z$ gate multi-controlled by $b_N$ qubits, requiring $4b_N$ T-gates and $b_N-1$ reusable qubits
    \item one $And$ gate, requiring $4$ T-gates and one reusable qubit.
\end{itemize}
The total complexity for the DF \sel circuit is:
\begin{itemize}
    \item    $24(L-1) + 28N(\beta-2) + 4L(10l_N + 4N+8\mu_N+7b_N-4+2k_N) + 8N + 4b_N + 4 = L(8 + 40l_N + 16N + 32\mu_N + 28b_N + 8k_N) + N(28\beta - 48) + 4b_N - 20$ T-gates
    \item  $6+b_L+2N+b_N+2\mu_N$ non-reusable qubits
    \item $2+\beta+b_N+b_L+\max[2\mu_N-1;b_N-1;l_N]+1+b_N+1+b_N-1+4=7+\beta+3b_N+b_L+\max[2\mu_N-1;b_N-1;l_N]$ reusable qubits
    \item $8L$ $R_Z$ rotations.
\end{itemize}
 
\subsubsection{MTD-$L^4$ decomposition}

\begin{figure*}
    \centering
    \includegraphics[scale=0.35]{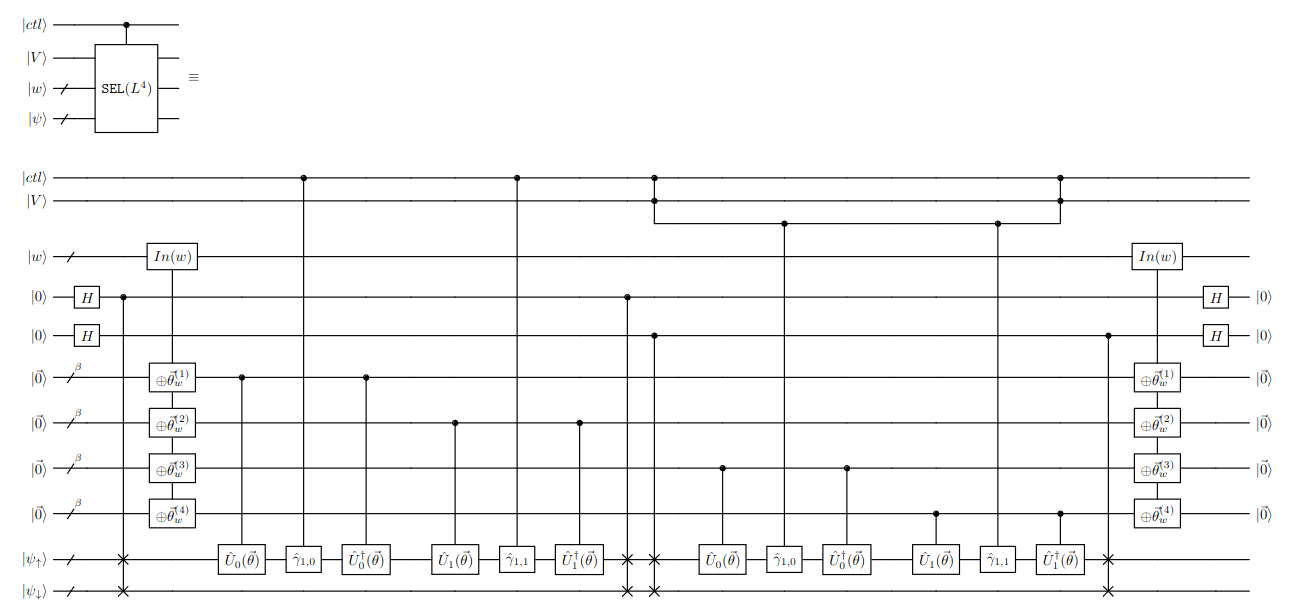}
    \caption{\sel circuit for implementing MTD-$L^4$ LCU.}
    \label{fig:sel_CP4T}
\end{figure*}

The implementation of the \sel circuit for the MTD-$L^4$ decomposition is shown in Fig.~\ref{fig:sel_CP4T}. This implementation minimizes the T-gate count in exchange for a higher number of ancilla qubits: four registers are used simultaneously for loading the Givens rotation angles, minimizing the number of times QROM must be called for angle loading. The cost for the MTD-$L^4$ \sel circuit is:
\begin{itemize}
    \item $4$ controlled swaps with an individual cost of $7N$ T-gates
    \item $2$ QROM circuits over $W$ indices, each one requiring $4(W-1)$ T-gates and $b_W - 1$ reusable qubits
    \item $8$ multiplexed Givens rotations circuits, each requiring $14N(\beta-2)$ T-gates
    \item one $And$ operation, requiring $4$ T-gates and one reusable qubit
    \item controlled single Majoranas, along with the Hadamard gates, which have no fault-tolerant cost.
\end{itemize}
The total complexity for the MTD-$L^4$ \sel circuit is:
\begin{itemize}
    \item $4*7N + 2*4(W-1) + 8*14N(\beta-2) + 4 = N(112\beta - 196) + 8W - 4$ T-gates
    \item $4 + 2N + b_W + 4\beta$ non-reusable qubits
    \item $b_W$ reusable qubits.
\end{itemize}

\subsubsection{Matrix Product States}
\begin{sidewaysfigure*}
    \centering
    \includegraphics[scale=0.35]{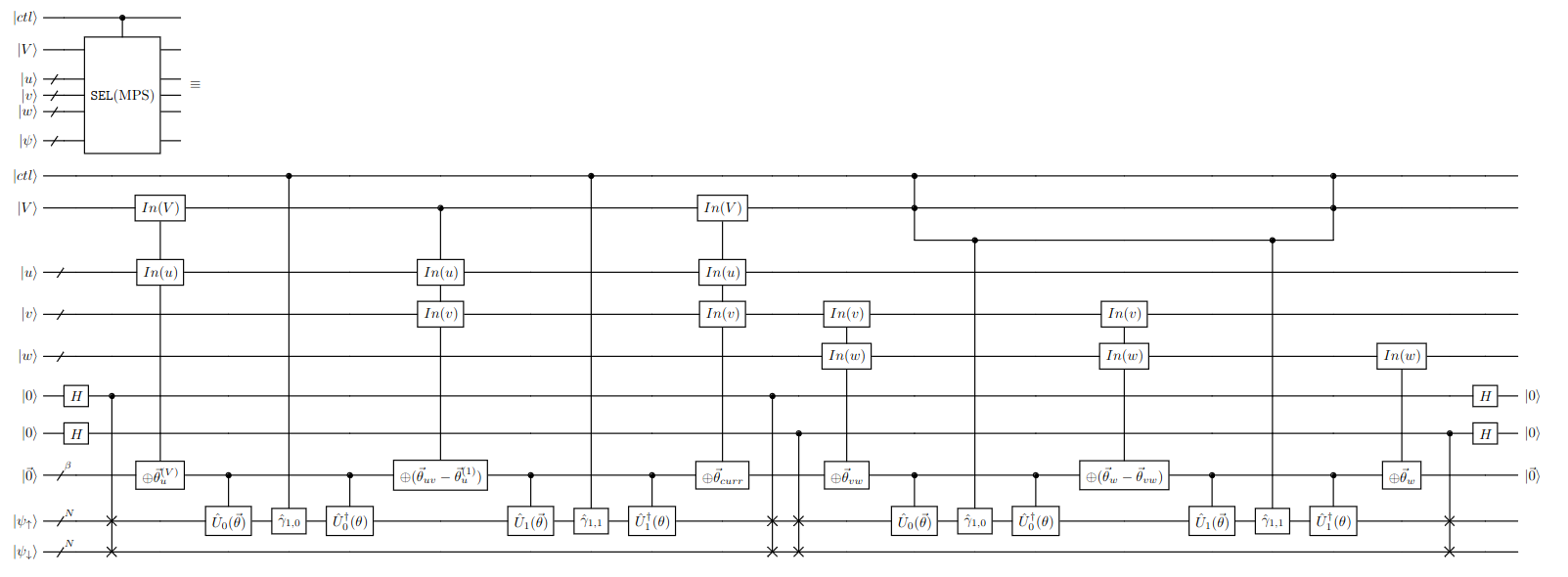}
    \caption{\sel circuit for implementing $L^4$-MPS LCU.}
    \label{fig:sel_MPS}
\end{sidewaysfigure*}

The \sel circuit for the $L^4$-MPS can be seen in Fig.~\ref{fig:sel_MPS}. The associated complexity components are:
\begin{itemize}
    \item one QROM over $N+\alpha_1$ indices, requiring $4(N+\alpha_1 - 1)$ T-gates and $b_{N+\alpha_1} - 1$ reusable qubits
    \item one controlled QROM over $\alpha_1\alpha_2$ indices, requiring $4(\alpha_1\alpha_2 - 1)$ T-gates and $b_{\alpha_1\alpha_2}$ reusable qubits
    \item one QROM over $(N+\alpha_1)\alpha_2$ indices, requiring $4((N+\alpha_1)\alpha_2 - 1)$ T-gates and $b_{(N+\alpha_1)\alpha_2} - 1$ reusable qubits
    \item two QROMs over $\alpha_2\alpha_3$ indices, each one requiring $4(\alpha_2\alpha_3 - 1)$ T-gates and $b_{\alpha_2\alpha_3} - 1$ reusable qubits
    \item one QROM over $\alpha_3$ indices, requiring $4(\alpha_3 - 1)$ T-gates and $b_{\alpha_3} - 1$ reusable qubits
    \item $4$ controlled swap gates, each one requiring $7N$ T-gates/
    \item $8$ multiplexed Givens rotations circuits, each requiring $14N(\beta-2)$ T-gates
    \item one $And$ gate for $4$ T-gates and one reusable qubit
    \item controlled single Majoranas and Hadamard gates for no additional fault-tolerant cost.
\end{itemize}
The total complexity for the MPS \sel circuit then becomes
\begin{itemize}
    \item $4(N+\alpha_1 + \alpha_1\alpha_2 + (N+\alpha_1)\alpha_2 + 2\alpha_2\alpha_3 + \alpha_3 - 6) + 4*7N + 8*14N(\beta-2) + 4 = N(4\alpha_2+ 112\beta - 192) + \alpha_2(8\alpha_1 + 8\alpha_3) + 8\alpha_1 + 4\alpha_3 - 24$ T-gates
    \item $4 + 2N + \beta + b_N + b_{\alpha_2} + b_{\alpha_3}$ non-reusable qubits
    \item $b_{\alpha_2\alpha_3}$ reusable qubits.
\end{itemize}

\section{Computational details}\label{app:compdet}

All electronic Hamiltonians in the minimal STO-3G basis set were obtained with the PySCF \cite{pyscf1,pyscf2,pyscf3} package. Results involving tensor decompositions of two-electron integrals correspond to the lowest rank sufficient to obtain a converged two-electron tensor with the 2-norm error $\sum_{ijkl}|\Delta g_{ijkl}|^2$ below $1\times 10^{-6}$. All LCU decompositions were done using the QuantumMAMBO package on Julia, which is available at \url{https://github.com/iloaiza/QuantumMAMBO.jl}.  All least square optimizations are performed using the Julia package \href{https://github.com/matthieugomez/LeastSquaresOptim.jl}{LeastSquaresOptim.jl}.

\subsection*{Molecular geometries}
For the molecules in Table~\ref{tab:results}, all bond lengths are taken to be $1 \textrm{\AA}$, $\angle$HOH = $107.6^\circ$ for H$_2$O, and $\angle$HNH = $107^\circ$ for NH$_3$.
For the hydrogen chains, the R(H -- H) distance was taken to be $1.4\textrm{\AA}$.

\subsection*{Details of THC implementation}

    The two-body tensor of the Hamiltonian $g_{i j k l}$, i.e., the terms of the Hamiltonian quartic in Majorana operators, for a system with $N$ spatial orbitals, is decomposed using 
    \begin{equation}
    g_{i j k l}=\sum_{\mu \nu} \Lambda_{\mu \nu} \left[\Vec{U}_\mu\otimes \Vec{U}_\mu^T\right]_{i j} \left[\Vec{U}_\nu\otimes \Vec{U}_\nu^T\right]_{k l} \label{thc_work_eq}
    \end{equation}
    where $\Lambda$'s are related to the $\zeta$'s from Eq.~\eqref{thc} as $\Lambda_{\mu \nu}=4\zeta_{\mu \nu}$ and $\vec{U}$'s are $N$ dimensional normalized vectors. \\
    To see that Eq.~\eqref{thc_work_eq} follows from Eq.~\eqref{thc}, one needs to rewrite Eq.~\eqref{thc} as
    \begin{equation}
        \hat{H} = \sum_{\mu\nu} \sum_{\sigma \tau} \zeta_{\mu\nu} \left(\hat{U}_\mu^{\dagger} (2\hat{n}_{1 \sigma}-1)\hat{U}_\mu\right)\left(\hat{U}_\nu^{\dagger} (2\hat{n}_{1 \tau}-1) \hat{U}_\nu\right) + \hat{h}_{1e}
    \end{equation}
    where $\hat{n}_{1 \sigma}$ and $\hat{n}_{1 \tau}$ are occupation number operators and $\hat{h}_{1e}$ contains only one-electron and constant terms. 
    Suppressing the spin-indices for clarity, we can rewrite the above expression as
    \begin{equation}
        \hat{H} = \sum_{\mu\nu} \Lambda_{\mu\nu} \left(\hat{U}_\mu^{\dagger} \hat{n}_{1}\hat{U}_\mu\right)\left(\hat{U}_\nu^{\dagger} \hat{n}_{1} \hat{U}_\nu\right) + \tilde{h}_{1e}, \label{thc:n+1e}
    \end{equation}
    where remaining constant terms have been absorbed into $\hat{h}_{1e}$ to give $\tilde{h}_{1e}$. 
    Since the one-electron terms can be treated optimally with DF, we focus on the first term of the RHS. We exploit the isomorphism between Lie algebra of one-electron excitation operators for a given spin and $N\times N$ matrix algebra, correspondingly, $\left(\hat{U}_\mu^{\dagger} \hat{n}_{1}\hat{U}_\mu\right)$ maps to $\mathbb{U}_\mu^{\dagger}\mathbb{n}_1\mathbb{U}_\mu$, where $\mathbb{U}_\mu$ and $\mathbb{n}_1$ are $N$-dimensional square matrices. The elements of $\mathbb{n}_1$ are given by $\left[\mathbb{n}_1\right]_{p q}=\delta_{p q}\delta_{p 1}$. It is straightforward to show that $\left[\mathbb{U}_\mu^{\dagger}\mathbb{n}_1\mathbb{U}_\mu\right]_{p q}=\left[\mathbb{U}_{\mu}\right]_{1 p}\left[\mathbb{U}_{\mu}\right]_{1 q}$. Since each row and column of a unitary matrix is a normalized vector, we can write $\left[\mathbb{U}_{\mu}\right]_{1 p}$ as the $p^{\rm th}$ component of a normalized vector $\Vec{U}$. 
    Thus, we can rewrite the first term in the RHS of Eq.~\eqref{thc:n+1e} as the RHS of Eq.~\eqref{thc_work_eq}.
    To arrive at the LHS of Eq.~\eqref{thc_work_eq}, the terms in the original Hamiltonian $\hat{H}$ quartic in fermionic creation and annihilation operators are considered. We note that the fermionic excitation operator $\hat{E}_q^p=\hat{a}_p^\dagger \hat{a}_q$ is isomorphic to the matrix $\mathbb{E}_q^p$ whose elements are $\left[\mathbb{E}_q^p\right]_{i j}=\delta_{i p}\delta_{j q}$. Mapping $\sum_{pqrs} g_{pqrs} \hat{E}_q^p \hat{E}_s^r$ to $\sum_{p q r s} g_{pqrs}\mathbb{E}_q^p \mathbb{E}_s^r$ and then exploiting the isomorphism we arrive at \eq{thc_work_eq}. The norm of the THC decomposition is $\sum_{\mu \nu} \abs{\Lambda_{\mu \nu}}$ in addition to $\Delta E/2$ of $\tilde{h}_{1e}$.
    
    For a system with $N$ spatial orbitals, we start with $K$ $\vec{U}$'s. All $\Lambda$'s are initialized to unity and each component of each $\vec{U}$ is initialized to $1/\sqrt{N}$. 
    We choose the target accuracy $\epsilon$ to be a residual 2-norm of $1\times10^{-6}$. The cost function for the THC decomposition is
\begin{equation}
    \sum_{p q r s} \left(g_{p q r s}-\sum_{\mu \nu} \Lambda_{\mu \nu} U_{\mu p} U_{\mu q} U_{\nu r} U_{\nu s}\right)^2+\sum_{\mu \nu}\left(\alpha\abs{\Lambda_{\mu \nu}}^\beta\right)^2
\end{equation}
where $\alpha$ and $\beta$ are hyper-parameters that tune relative importance of the two contributions. 
The first term ensures the convergence of the THC ansatz to the original two-body tensor, while the second term penalizes
large 1-norm. We found that the optimal decompositions are obtained via the choice of parameters $\alpha=0.2$ and $\beta=1$.
 The form of the cost function allows us to optimize for $\Lambda$'s and $\Vec{U}$'s via non-linear least square optimization, which we carried out using the Levenberg-Marquardt algorithm~\cite{Levenberg,Marquardt}.
 If the 2-norm of the residual two-body tensor is greater than the desired accuracy $\epsilon$, we decompose the residual two-body tensor using another set of $K$ normalized vectors. 
 The initialization of the new $\Lambda$'s and $\Vec{U}$'s is the same as before. This step is repeated until the desired accuracy is achieved. 
 A total of $J$ sets each containing $K$ normalized vectors are required to achieve the desired accuracy. We worked with $K=3N$ for the H-chains reported in Fig.~\ref{fig:linear} and Table~\ref{tab:fits}, and $K=4N$ for the small molecules reported in Table~\ref{tab:results}, with the exception of NH$_3$ for which $K=36$. For all THC decompositions reported in this paper, we found that $J=2$ was sufficient to reach the target accuracy. In order to estimate the resources required to encode the THC decomposition into a quantum circuit, we have assumed that the allowed error in the quantum phase estimation is 1 mili-Hartree and the number of bits used to represent the LCU coefficients and rotation angles are 10 and 16, respectively.



%

\end{document}